\documentclass[aps,
pra,
twocolumn,
floats,
epsfig,
showpacs]{revtex4-1}

\usepackage[apple mac]{inputenc}
\usepackage{textcomp}

\usepackage{bbm}
\usepackage{amssymb}
\usepackage{ifpdf}
\usepackage{color}
\usepackage{graphicx}
\usepackage{amsmath}
\usepackage{amsfonts}
\usepackage{hyperref}
\usepackage{bbold}
\usepackage{bm}
\usepackage{color}
\usepackage{verbatim}

\newcommand{\qed}{\hspace*{\fill}$\square$}

\newcommand{\be}{\begin{equation}}
\newcommand{\ee}{\end{equation}}
 \newcommand{\ket}[1]{|#1\rangle}
 \newcommand{\bra}[1]{\langle #1|}

\begin{document}

\title[Short Title]{Quantum Google in a Complex Network}

\author{G.D. Paparo$^{1}$, M. M\"uller$^{1}$, F. Comellas$^{2}$ and M. A. Martin-Delgado$^{1}$}
\affiliation{$^1$Departamento de Fisica Teorica I, Universidad Complutense, 28040 Madrid, Spain. \\
$^2$Departament Matem\`atica Aplicada IV, 
     Universitat Polit\`ecnica de Catalunya, 08034 Barcelona,
     Spain.}
\begin{abstract}
We investigate the behavior of the recently proposed quantum Google algorithm, or quantum PageRank, in large complex networks. Applying the quantum algorithm to a part of the real World Wide Web, we find that the algorithm is able to univocally reveal the underlying scale-free topology of the network and to clearly identify and order the most relevant nodes (hubs) of the graph according to their importance in the network structure.
Moreover, our results show that the quantum PageRank algorithm generically leads to changes in the hierarchy of nodes.  In addition, as compared to its classical counterpart, the quantum algorithm is capable to clearly highlight the structure of secondary hubs of the network, and to partially resolve the degeneracy in importance of the low lying part of the list of rankings, which represents a typical shortcoming of the classical PageRank algorithm. Complementary to this study, our analysis shows that the algorithm is able to clearly distinguish scale-free networks from other widespread and important classes of complex networks, such as Erd\H{o}s-R\'enyi networks and hierarchical graphs. We show that the ranking capabilities of the quantum PageRank algorithm are related to an increased stability with respect to a variation of the damping parameter $\alpha$ that appears in the Google algorithm, and to a more clearly pronounced power-law behavior in the distribution of importance among the nodes, as compared to the classical algorithm. Finally, we study to which extent the increased sensitivity of the quantum algorithm persists under coordinated attacks of the most important nodes in scale-free and Erd\H{o}s-R\'enyi random graphs.
\end{abstract}

\maketitle

%%%%%%%%%%%%%%%%%%%%%%%%%%%%
%%%%%%%%%%%%%%%%%%%%%%%%%%%%
\section{Introduction}
\label{sect_intro}
%%%%%%%%%%%%%%%%%%%%%%%%%%%%
%%%%%%%%%%%%%%%%%%%%%%%%%%%%

It is of great interest to explore and classify the large amount of information that is stored in huge complex networks like the World Wide Web (WWW).
A central problem of bringing order to classical information stored in networks such as the WWW amounts to rank nodes containing such information according to their relevance. A highly successful and nowadays widespread tool for this purpose has been the PageRank algorithm~\cite{brin1998anatomy,page1999pagerank}, which lies at the core of Google's ranking engine. In the foreseeable future where large-scale quantum networks have become a reality, classifying the quantum information stored therein will become a priority. It is in this sense that the recently introduced quantum PageRank algorithm \cite{gdpmamd2011} is an important achievement as it constitutes a quantization of the classical PageRank protocol. This new quantum algorithm has shown, applied to small networks, a striking behavior with respect to its classical counterpart, such as producing a different hierarchy of nodes together, paired with a better performance. In this paper we investigate the properties of the quantum algorithm for networks which model large real-world complex systems. We also test the algorithm on real-world data stemming from a part of the WWW.

Complex networks are more and more pervasive and essential in our everyday's life, and thus 
the importance of network science. 
Consequently, considerable research is devoted  to analyze and understand networks 
like the World Wide Web, the Internet, networks associated to  transportation and communication
systems and even  biological and social networks.
Starting with the seminal papers by Watts and
Strogatz on small-world networks~\cite{small-world} and by Barab\'asi and
Albert on scale-free networks~\cite{Barabasi_Albert_99}, researchers realized that most relevant networks belong to a class  known as small-world scale-free networks, as they exhibit both strong local clustering  (nodes have many mutual neighbors) and a small average path length while sharing another important characteristic: the number of links of nodes usually obeys a power-law distribution (the network is scale-free). Moreover, it has  been found that many real networks, including the WWW,  are also self-similar, see \cite{Song_2005}.
Such properties can often be related to a modular and hierarchical structure and organization which is 
essential for their communication and dynamical processes~\cite{RaBa03,Barabasi-Oltvai-04,BaDeRaYoOl03}. 
On the other hand, this hierarchical structure could explain the existence of nodes 
with a relatively large number of links (or hubs), which play a critical role in the information flow of the system. Hubs are also associated with a low average distance in the network. Several general reviews and books on complex networks  are now available, to which we refer the reader interested in more information on this topic~\cite{statistical2,Newman-Nets,CoHa10}. 

Networks considered in this paper are modeled by three classes of graphs:  The first type are Erd\H{o}s-R\a'enyi random graphs~\cite{Erdos_Renyi_1959} . These graphs are constructed by connecting a given set of nodes with directed edges, each one added according to a certain fixed probability. 
The second class of graphs are scale-free graphs which were introduced by Barab\'asi and Albert to model the WWW~\cite{Barabasi_Albert_99,Barabasi_13}. A graph is dynamically formed by a continuous addition of new vertices which are connected preferentially to vertices which already have a large degree. We consider here a version for directed graphs published in~\cite{BoBoChRi03}. The third family are hierarchical graphs which are also scale-free but their clustering and degree distributions are negatively correlated as hubs have a smaller clustering coefficient than nodes with a lower degree. It has been found that hierarchical graphs constitute also a good model for the WWW ~\cite{RaBa03,Barabasi-Oltvai-04}.

In the next sections we apply the quantum Google algorithm proposed in ref.~\cite{gdpmamd2011} to representatives of the three classes of graphs. 
Our focus lies on directed scale-free networks, and also on hierarchical graphs as they are  good models for the WWW, but we in addition also consider Erd\H{o}s -R\'enyi random networks as a reference in order to contrast the results which we find for the former models. 

In particular, two main fundamental questions will be addressed:

1/ Does the quantum PageRank algorithm preserve the structure of a scale-free network? In other words, we ask whether the ranking distribution obtained by the quantum algorithm also follows the same pattern of node-importance as the underlying scale-free network, and does not get mixed up with the distribution corresponding to a random Erd\H{o}s-R\a'enyi network.
Similarly, for a  Erd\H{o}s-R\a'enyi
network we study whether the behavior of the quantum Page Rank algorithm is intrinsic to these networks and to which extent it differs from the one for scale-free networks. 

2/ Can one improve on the information gained on some of the
properties of a scale-free network as far as the quantum Google  algorithm is
concerned? Specifically, scale-free networks are robust against
random disturbances which are unavoidable in a noisy
environment. However, scale-free networks are vulnerable to coordinated
attacks.

These questions are outlined in more detail throughout  the paper and clear answers are given. We hereby summarize briefly some of our main results:

i/ We find that the quantum PageRank algorithm detects each of the representatives classes of complex networks, in the sense that 
the classification results show that the ranking of the 
 nodes follow a similar law than the degrees of the network. 
 In particular, for a scale-free network this distribution remains scale-free under the application of the quantum PageRank algorithm. This implies that the quantum PageRank  algorithm is expected to be robust with respect to random external noise.

ii/ Remarkably enough, the detection of the hubs for a network in the class of scale-free networks with the QG algorithm is clearly enhanced with respect to the classical PageRank algorithm. In particular, the quantum PageRank algorithm is able to more clearly reveal the existence of  secondary hubs in scale-free networks.

iii/ We find that the quantum PageRank algorithm and the resulting rankings are more stable than the classical PageRank protocol with respect to the variation of the damping parameter 
$\alpha$ that appears in the Google algorithm. The low dependence on this parameter, which inevitably had to be arbitrarily tuned in the classical algorithm, provides the quantum algorithm with a higher objectivity of the importance rankings.

iv/ Our study shows that the quantum PageRank algorithm displays for scale-free graphs a power law scaling behavior of the importances of the nodes. Furthermore, this power law behavior is more favorable than the one the classical algorithm exhibits: Indeed, a smoother behavior is connected to a more harmonious distribution of importance among the nodes, which enables the algorithm to better uncover the structure of hubs in the underlying scale-free graphs.

v/  The enhanced sensitivity of the quantum PageRank algorithm to structural details of the networks is related to an increased sensitivity of the quantum algorithm under coordinated attacks of the most important nodes in scale-free networks.

%%%%%%%%%%%%%%%%%%%%%%%%%%%%%%%%%%%%%%%%%
%%%%%%%%%%%%%%%%%%%%%%%%%%%%%%%%%%%%%%%%%
\subsection{Conceptual Setting of the Quantum PageRank Algorithm}
\label{subsect_intro_0}
%%%%%%%%%%%%%%%%%%%%%%%%%%%%%%%%%%%%%%%%%
%%%%%%%%%%%%%%%%%%%%%%%%%%%%%%%%%%%%%%%%%

Motivated by the fact that in a near-future scenario a certain class of quantum network will be operative~\cite{darpa,SECOQC,UQCC,SwissQuantum,vicente1,ETSI}, but not yet a scalable quantum computer, in~\cite{gdpmamd2011} a class of quantum algorithms  to rank the nodes in a quantum network was put forward. The algorithms in this class must be compatible with the classical one. Indeed, existing projects for large-scale quantum networks contemplate using the backbone of existing communication networks, upgrading them to include the quantum hardware in order to store and manipulate quantum information.
In particular, the directed structure  of the graph must be preserved, a feature which is crucial to measure a node's  authority. An instance of this class  of algorithms was explicitly constructed and analyzed for graphs of small size and it was found to serve as a valid quantum counterpart of Google's PageRank algorithm (for details see ref.~\cite{gdpmamd2011}).

To perform such a \emph{quantum task}, it is important that the ranking algorithm must in principle incorporate some of the quantum properties of the network, like quantum fluctuations, and be objective. Indeed, the latter property earned much of Google's PageRank's success and was achieved embedding in the algorithm the random walk of a surfer exploring the WWW based on simple sensible rules. 
Within the same line of reasoning, we set up a quantum walk based algorithm that mimics the exploration of nodes in a quantum network, in the setting when these are represented by states of a Hilbert space. The simple rules are encoded in the quantum dynamics, and in doing so the quantum nature of the networks and the information stored is properly taken into account.
In the setting where a fully fledged large-scale quantum computer is not yet available, a key property of the quantum algorithm is that it be efficiently simulatable on a classical computer, that is, it must belong to the computational complexity class P. Our algorithm is based on a single particle quantum walk and thus efficiently simulatable.

Furthermore, the fact that the quantum algorithm contains a quantum walk at its heart paves the way to the analysis of its dynamics from a purely physical perspective: In our work, we analyze extensively the localization properties of the quantum walk contained in the algorithm applied to several classes of networks (see sect. IV) showing its effects on ranking. To do so, we introduce the new concept of a quantum Inverse Participation Ratio (IPR). This is a generalization of the classical IPR, which is the main quantity that has been used extensively to probe localization properties in the study of classical random walks.

Besides its application to future quantum networks, the Quantum PageRank algorithm can indeed also be regarded as a valuable quantum tool, that can be efficiently run on a a classical computer, to perform a "classical" task, namely to rank nodes in a classical networks. Indeed, embedding nontrivially the network connectivity structure in the quantum dynamics, our protocol turns out to show several features that improve those present in the classical algorithm.

Here we do not concentrate on possible quantum speedups and a detailed resource analysis for the quantum algorithm, which lies in the computational complexity P of efficiently simulatable algorithms. 
Instead, in this work we focus on the advantages that ranking nodes in classical networks using quantum algorithms displays, such as an increased resolution in the structure analysis of scale-free graphs or an increased stability with respect to the variation of the damping parameter.

%%%%%%%%%%%%%%%%%%%%%%%%%%%%%%%%%%%%%%%%%
%%%%%%%%%%%%%%%%%%%%%%%%%%%%%%%%%%%%%%%%%
\subsection{Operational Summary of the Quantum PageRank Algorithm}
\label{subsect_intro}
%%%%%%%%%%%%%%%%%%%%%%%%%%%%%%%%%%%%%%%%%
%%%%%%%%%%%%%%%%%%%%%%%%%%%%%%%%%%%%%%%%%

\begin{figure*}
\centering
\includegraphics[keepaspectratio=true,width=.93\linewidth]{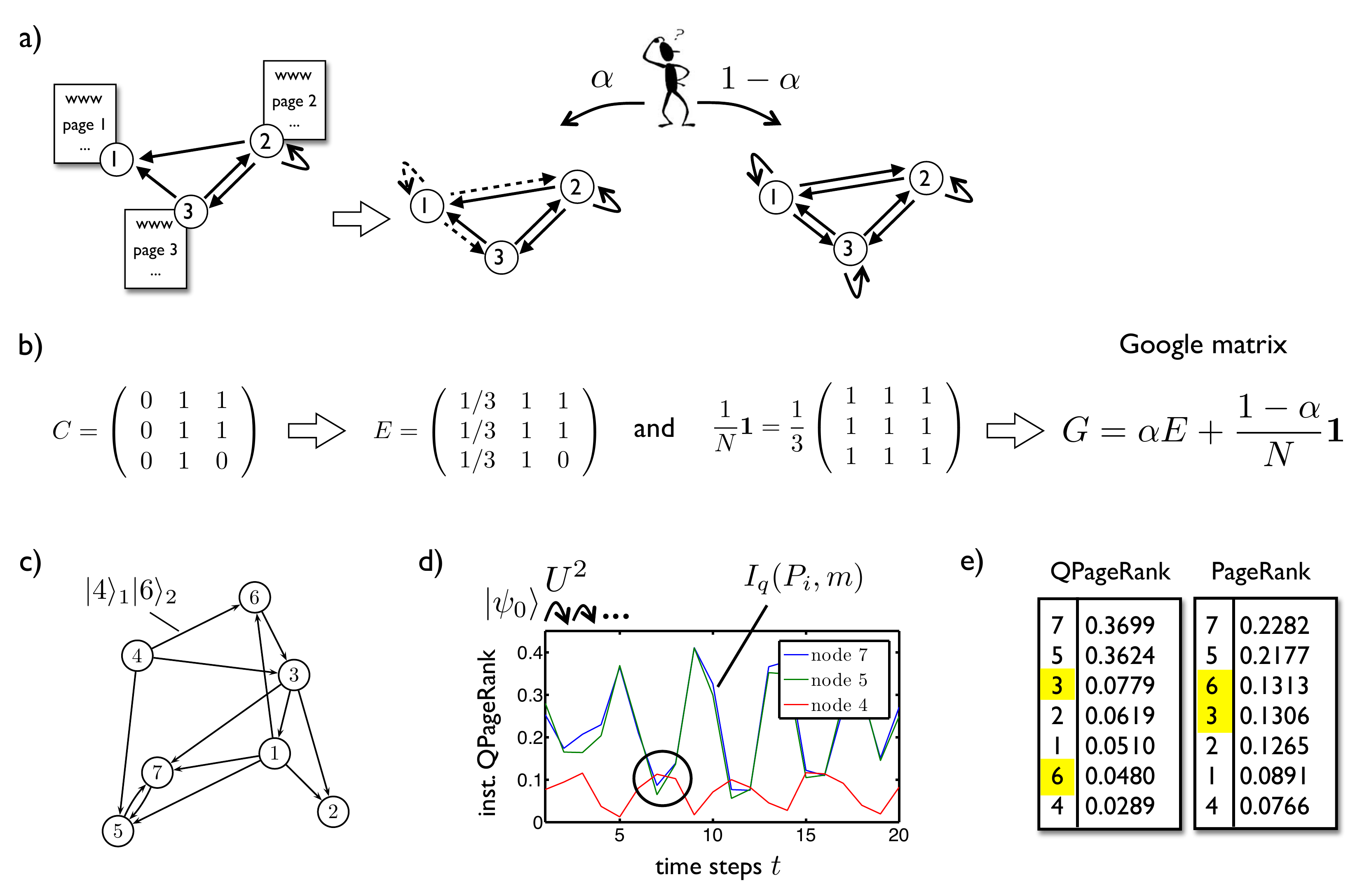}
\caption{
\label{fig:QGoogleSchematics}
(Color online) Schematic outline and summary of the quantum PageRank algorithm as proposed in \cite{gdpmamd2011}.
a) The internet can be thought of as a set of pages (nodes of a graph) connected by directed hyperlinks (edges of the graph). The classical (quantum) PageRank algorithm can be regarded as a single walker performing a directed classical (quantum) random walk on the graph. 
b) The connectivity structure of the graph, as described by the connectivity matrix $C$, is of paramount importance to perform the ranking of the importance of pages both in the classical and the quantum case. In the classical (quantum) case the walker performs an incoherent  (coherent) walk according to a combination of two hopping processes along the graph. The dynamics is governed by the "Google matrix" $G$, which describes the dynamics as a combination of (i) hopping according to a (patched) connectivity matrix $E$ of the graph (parameter $\alpha$, see main text and \cite{gdpmamd2011} for more details) and (ii) a fully random hopping processes (parameter $1-\alpha$), where each node is connected to all other nodes of the graph. 
c) In the quantum PageRank algorithm the Hilbert space is spanned by the set of directed links between all pairs $i$ and $j$ of nodes of the graph, as tensor product states $\ket{i}_1 \ket{j}_2$. The initial state $\ket{\psi_0}$ as well as the coherent discrete time evolution operator $U^2$ for the directed quantum walk (see main text) are determined by the Google matrix $G$. 
d) Quantum fluctuations can lead to a reversal of the order of importances of pages at certain instances of (discrete) time (so-called instantaneous outperformance), as well as on the average over longer times (average outperformance). The latter effect is reflected by changes in the ordered list of pages (nodes) when the importance of pages according to the quantum PageRank algorithm is compared to its classical counterpart, the list of classical PageRank values. Red, blue and green curves in d) show the instantaneous quantum PageRank of nodes \# 4, 5 and 7 of the seven-node-graph shown in c), which was explored in Ref.~\cite{gdpmamd2011}, and leads to the quantum PageRank and classical PageRank lists displayed in e).}
\end{figure*}

In this section we briefly review the quantum PageRank algorithm from an operational point of view, whereas the reader can find the details of the construction in \cite{gdpmamd2011}. This quantum PageRank algorithm  
satisfies all the properties of this class and represents a valid quantization of Google's PageRank algorithm. A step-by-step illustration of the quantum algorithm is
presented in Fig.~\ref{fig:QGoogleSchematics}.

\noindent In Google PageRank's algorithm the ranking is performed setting up a random walk on the network. The walk uses as a transition matrix, known as the Google matrix $G$. The Google matrix is the weighted sum of two transition matrices. The first walk is driven by a modified connectivity matrix $E$ where outgoing links to all other nodes have been added to every node that has no outgoing link. The second walk is a simple random hopping that connects every node to any other node. 
Accordingly, the Google matrix associated to a given graph is defined as follows:
\begin{equation}
G := \alpha E + \frac{(1- \alpha )}{N} \mathbf{1},
\label{eq:_google_matrix_pagerank}
\end{equation}
\noindent  where $\mathbf{1}$ is a matrix with entries all set to $1$ and $N$ is the number of nodes. The parameter $\alpha$ is known as the \emph{damping} parameter.

 \noindent The ranking of the nodes is performed measuring the probability to find the walker on each node when the stationary distribution $I$ has been reached i.e. when 
$G I = I$.
%
%%
%\begin{twocolumn}
%\end{twocolumn}
%%

In the quantum PageRank algorithm the idea is to set up a quantum walk on the nodes of the network and to perform the ranking of the nodes measuring the probability of finding the quantum walker on such nodes.

\noindent  The quantum walk is a quantization of the Markov chain underlying the classical PageRank algorithm and is set up using Szegedy's method  \cite{Szegedy04} which allows one to take into account explicitly the connectivity structure, and the directedness of the network in particular. 
One can quantize a Markov chain on a $N$-vertex graph that has as transition matrix, the Google matrix $G$. This is performed by introducing a discrete-time quantum walk embedding the stochastic $N \times N$ matrix $G$ on the same graph.

The Hilbert space is the span of all vectors representing the $N \times N$ (directed) edges  of the graphs i.e. ${ \mathcal H }=  {\rm span}\{ | i\rangle_1 | j \rangle_2 \, ,{\rm with}\, i , j \in N\times N\} = {\mathbb C}^{N} \otimes {\mathbb C}^{N}$, where the order of the spaces in the tensor product is meaningful here because we are dealing with a directed graph. 

\noindent For each vertex let us define the quantum state vector,
\begin{equation}
| \psi_j \rangle :=    \sum_{k=1}^{N}  \sqrt{ G_{kj} } \, |  j \rangle_1  \,| k \rangle_2\quad , \forall j 
\label{ eq: Sgezedy vector 1}
\end{equation} 
\noindent that is a superposition of the quantum states representing the edges outgoing from the $j^{th}$ vertex. The weights are given by the (square root of the) Google matrix $G$.

The idea of the quantum PageRank satisfying all the properties that define the class of possible ranking algorithms, is to set up a quantum walk starting from the initial state  $| \psi_0 \rangle = \frac{1}{\sqrt{N}} \sum_{i=1}^{N} | \psi_j\rangle $ and whose dynamics is governed by the quantum evolution operator $ U := S (2 \Pi -\mathbb{1}) $ where $S$ is the swap operator i.e. $S=\sum_{j,k=1}^{N} \ket{j} \ket{k} \bra{k} \bra{j} $
and  $\Pi := \sum_{j=1 }^{N} | \psi_j \rangle  \langle  \psi_j |,$

\noindent Although the Hilbert space is clearly $N^2$ dimensional it can be shown that the dynamics takes place in an invariant subspace which is at most $2N$ dimensional. This allows one to numerically treat networks with a larger number of nodes.

\noindent The rankings of the nodes in the quantum network is calculated using the \emph{instantaneous quantum PageRank}:
\begin{equation}
I_q(P_i,t) =  \langle  \psi_0 | \,  {U^\dagger}^{2t} | i \rangle_2  \langle  i |  U^{2t} | \psi_0 \rangle.
\label{ eq: Quantization of PR importance}
\end{equation} 
\noindent Another quantity, called the  \emph{average quantum PageRank} can be defined:
\begin{equation}
\langle I_q (P_i)\rangle:=\frac{1}{T} \sum_{t=0}^{T-1}  \ I_q (P_i,t).
\label{eq:average_QPR_def}
\end{equation}
Whereas this quantity can be shown to converge for $T$ sufficiently large, the \emph{Instantaneous Quantum PageRank} does not converge in time \cite{Aharonov2002}.

\noindent In \cite{gdpmamd2011} nontrivial features of the quantum PageRank were uncovered, such as the \emph{instantaneous outperformance} of the algorithm or the violation of the hierarchy as predicted by the classical algorithm.

These properties motivate us to investigate the persistence of these novel effects on larger complex networks. 
This paper is organized as follows: 
In section \ref{sect_II} we present an analysis on graphs of the scale-free type for networks with hundreds of nodes, and in particular for a real-world network.
In \ref{sect_III})  we have extended the analysis of the quantum PageRank algorithm on Erd\"os-R\'enyi and hierarchical graphs. From this we conclude that the behavior
of the quantum PageRank is characteristic of each type of complex networks, and moreover, it is different from the classical PR algorithm: the classical ranking changes by quantum fluctuations.
The new quantum dynamics incorporated in the task of ranking also raises other important questions on the properties of the quantum walk embedded in the algorithm such as the localization phenomena of the walker on the network is analyzed in section \ref{sect_IV}, while  the stability of the ranking with respect to noise (or {\it damping} parameter)  is studied in section \ref{sect_V}. 
In section  \ref{sect_VI} we address the  question of whether the power law behavior displayed by the classical PageRanks is preserved by the quantized algorithm and compute the scaling exponent. In section \ref{sect_VII} se investigate a very practical situation and study how sensitive the quantum PageRank algorithm is under coordinated attacks in scale-free graphs. Section \ref{sect_conclusions} is devoted to conclusions and outlook, and an appendix provides some technical details.

%%%%%%%%%%%%%%%%%%%%%%%%%%%%%%%%%%%%%%%%%
%%%%%%%%%%%%%%%%%%%%%%%%%%%%%%%%%%%%%%%%%
\section{Quantum PageRank on Scale-Free Networks}
\label{sect_II}
%%%%%%%%%%%%%%%%%%%%%%%%%%%%%%%%%%%%%%%%%
%%%%%%%%%%%%%%%%%%%%%%%%%%%%%%%%%%%%%%%%%
We will analyze the predictions of  the quantum PageRank algorithm on complex networks. 
We will focus on random scale-free networks because of their widespread appearance and relevance in real-world applications. In the next section we will also deal with the important cases of random (Erd\"os - R\'enyi) and hierarchical networks for completeness in order to check whether the quantum PageRank  algorithm preserves the characteristics of different classes of complex networks. Moreover, the study of random and hierarchical networks results to be useful to confront its features with scale-free networks and to draw interesting conclusions.

Random scale-free graphs \cite{Barabasi_RMP_2002,Boccaletti_2006} are ubiquitous in nature. They appear as good  models of the World Wide Web \cite{Barabasi_Albert_Jeong_2000}, airline networks \cite{Barrat_2004} or metabolic networks \cite{Jeong_Mason_2001,Jeong_Tombor_2000}, just to name a few. These are networks that display a small fraction of hubs, i.e. nodes with a high connectivity, a property that follows from the degree distribution $P(k)$ that shows a scale-free behavior, $P(k) \approx k^{-\gamma }$. Scale-free networks exhibit intriguing properties which have been studied extensively, such as robustness against uncoordinated attacks~\cite{Albert_Jeong_Barabasi_2000,Callaway_Newman_2000,Vazquez_Moreno_2003}, good navigability \cite{Boguna_2009, Carmi_Carter_2009,Lee_Holme_2000} and controllability \cite{Liu_Slotine_2000,Nepusz_2012,Nicosia_Latora_2012}.

One of the first models proposed to describe scale-free networks is the \emph{preferential attachment model}~\cite{Barabasi_Albert_99,Barabasi_13}. 
In this model links are preferentially formed to already highly connected nodes.
A random directed scale-free model for the WWW was also introduced in~\cite{diameterWWW} and a generalization 
appeared in~\cite{BoBoChRi03}. To produce the characteristic  power-law degree distribution of degrees, the models consider 
two main mechanisms: growth and preferential attachment.
A graph is dynamically formed by a continuous addition of new vertices and 
each new vertex is joined to several existing vertices selected proportionally  to
their in and out degrees. The generalized model allows also the introduction of directed 
edges between two already existing nodes.
In this work we use  graphs created  with this  model as implemented in NetworkX~\cite{HaScSw08}.

\begin{figure}
\includegraphics[keepaspectratio=true,width=.95\linewidth]{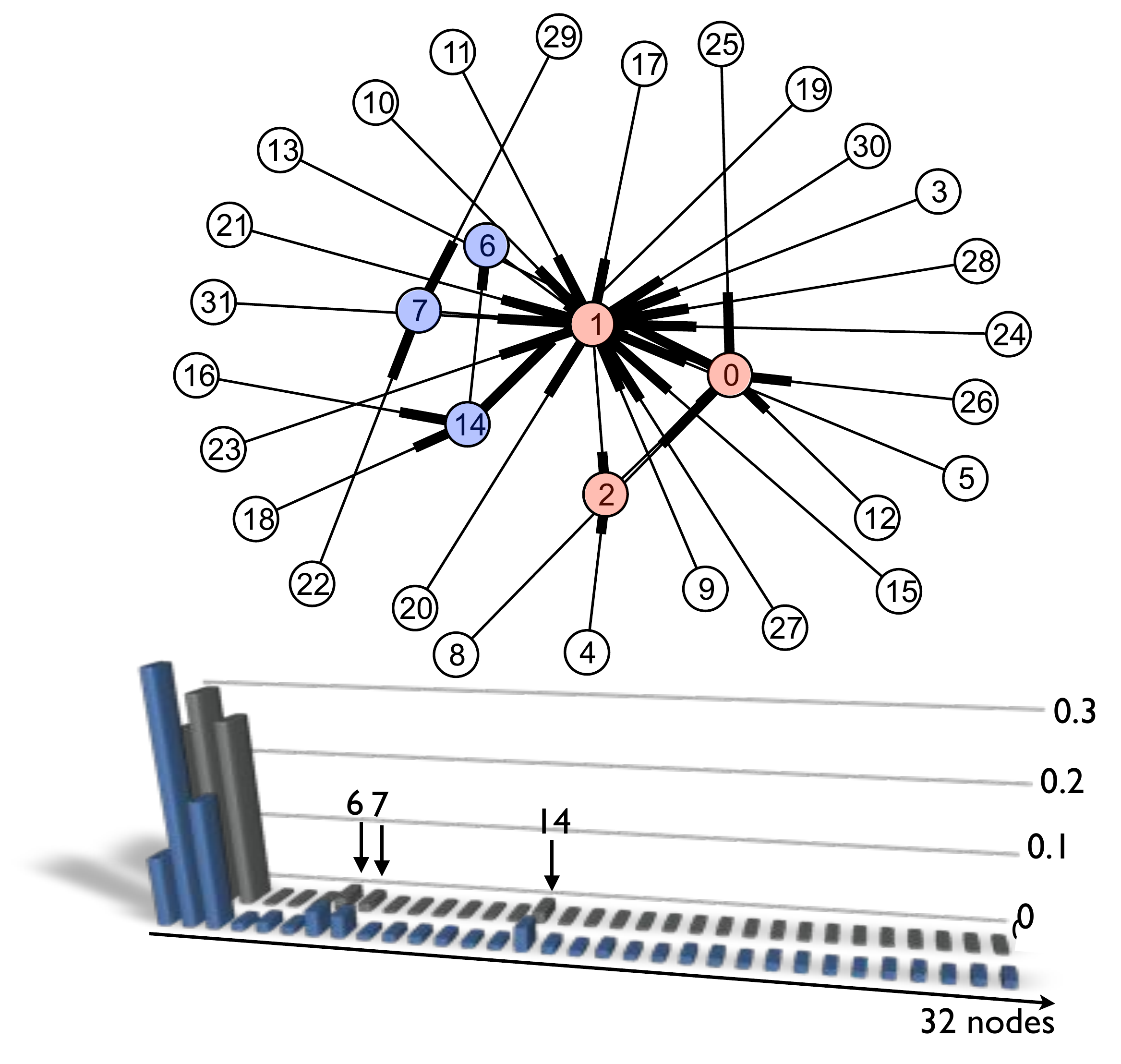}
\caption{
\label{fig:SF_32_nodes_Graph_layout}
(Color online) Scale-free graph with 32 nodes and the comparison of the importance of the nodes when evaluated with the quantum and classical PageRank. The classical PageRank shows a very sharp concentration of importance on the three main hubs, nodes 0, 1 and 2. One can see from the comparison of the predictions of the two algorithms the relative emergence of secondary hubs (nodes 6, 7 and 14) when the importance is calculated with the quantum PageRank (see text in sect. \ref{sect_II})
}
\end{figure}

\begin{figure}
\includegraphics[keepaspectratio=true,width=.95\linewidth]{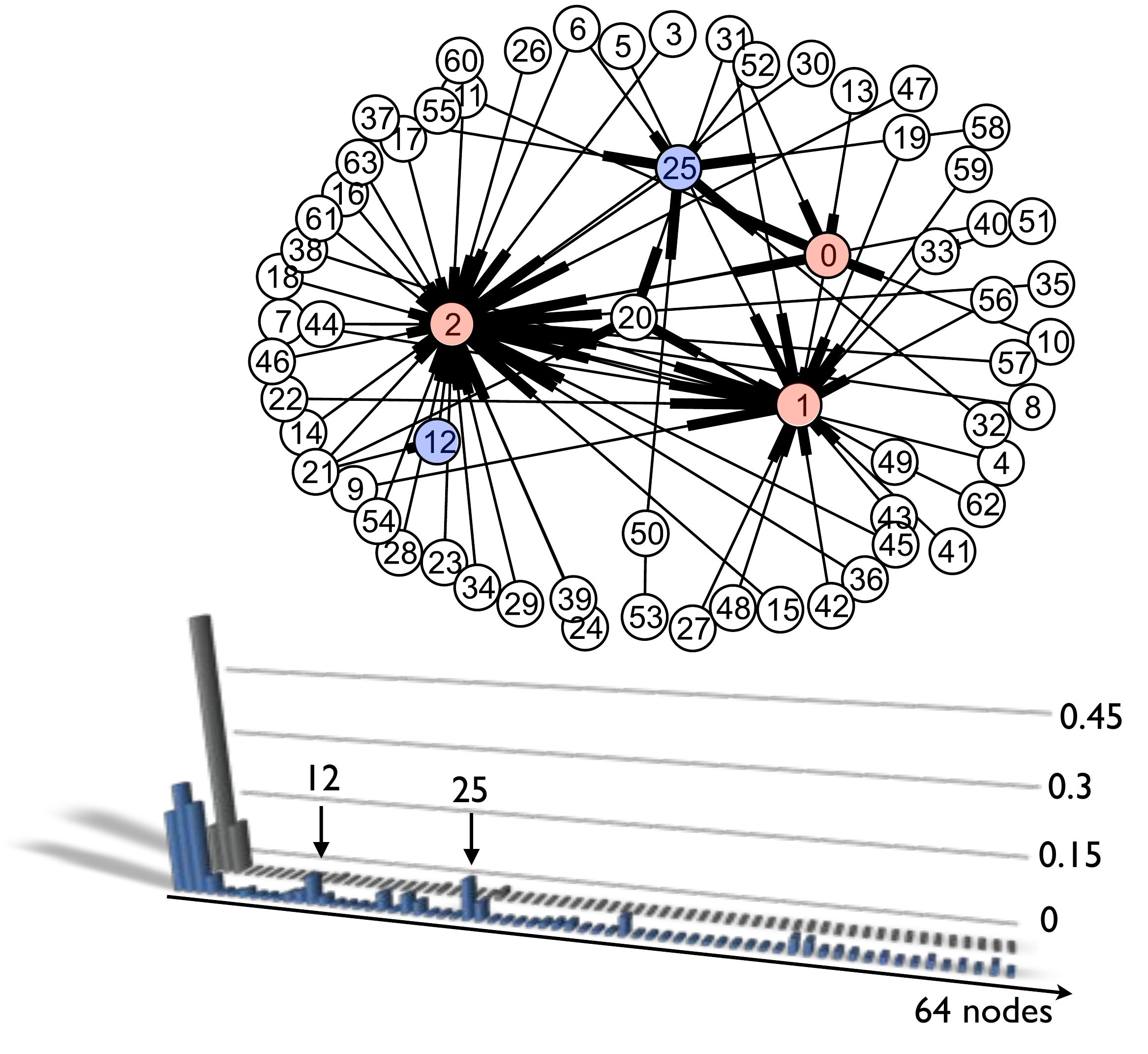}
\caption{
\label{fig:SF_64_nodes_Graph_layout}
(Color online) Scale-free graph with 64 nodes and the comparison of the importance of the nodes when evaluated with the quantum and classical PageRank. The classical PageRank shows a very sharp concentration of importance on the three main hubs, nodes 0, 1 and 2. The quantum PageRank algorithm is able to better distinguish and to highlight the secondary hubs (in this graph nodes: 12 and 25) whose importance rises with respect to the primary hubs. (see text in sect. \ref{sect_II})
}
\end{figure}

\begin{figure}
\includegraphics[keepaspectratio=true,width=.95\linewidth]{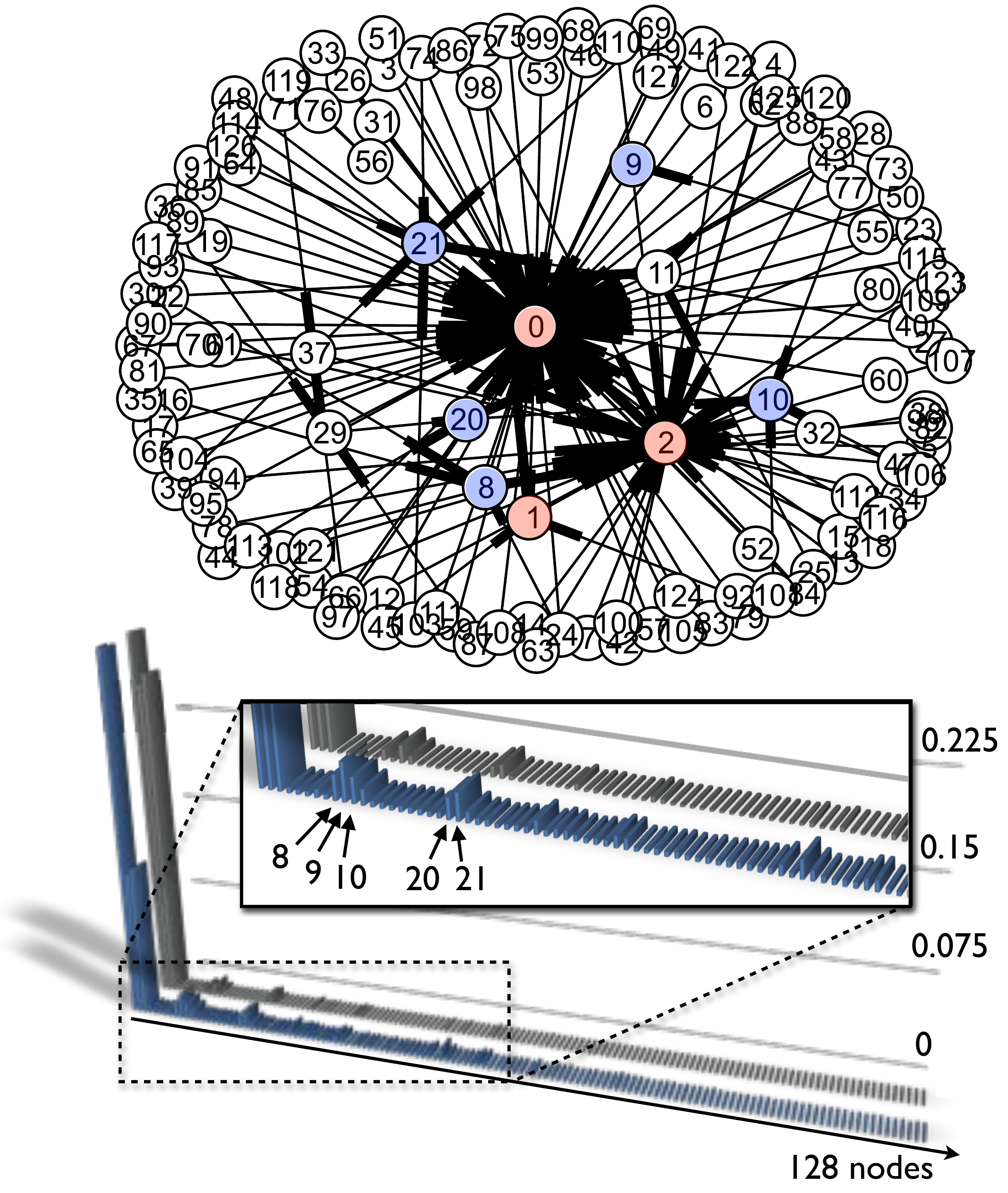}
\caption{
\label{fig:SF_128_nodes_Class_PR_vs_QPR}
(Color online) Scale-free graph with 128 nodes (see text in sect. \ref{sect_II})
and a comparison of the importance of the nodes when evaluated with the quantum and classical PageRank. The classical PageRank shows a very sharp concentration of importance on the three main hubs, nodes 0, 1 and 2. One can see from the comparison of the predictions of the two algorithms the relative emergence of secondary hubs (nodes 8, 9, 10, 20 and 21) when the importance is calculated with the quantum PageRank (see text in sect. \ref{sect_II}).
}
\end{figure}

We next discuss results originating from the application of the quantum PageRank to complex networks of the scale-free type with hundreds of nodes (${\mathcal O}(100)$). 
To this end, we have performed a numerical analysis using the quantum PageRank algorithm. We find that the algorithm clearly identifies that the networks are of scale-free type. The algorithm is able to point out the most important hubs. This is a task already well performed by the classical PageRank. However, the quantum PageRank algorithm has improved ranking capabilities, not concentrating all the importance on these few nodes. Indeed, it is capable of unveiling the structure of the graph highlighting also the secondary hubs (see figures \ref{fig:SF_32_nodes_Graph_layout} , \ref{fig:SF_64_nodes_Graph_layout}, \ref{fig:SF_128_nodes_Class_PR_vs_QPR} and captions therein).

\begin{figure}
\includegraphics[keepaspectratio=true,width=.96\linewidth]{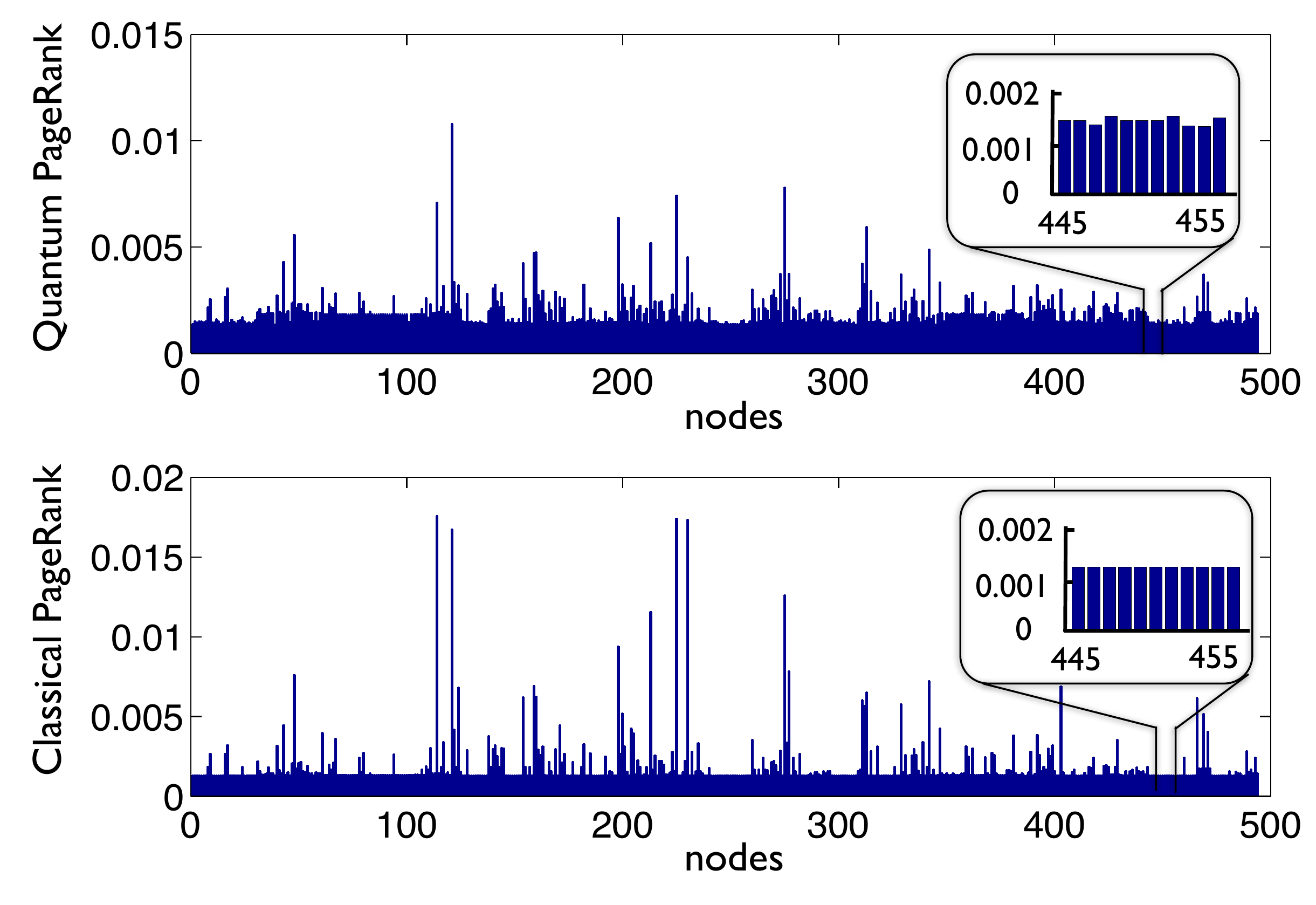}
\caption{
\label{fig:MM_EPA_900_2300_Class_PR_vs_QPR}
(Color online) Comparison of the quantum and classical PageRank on a real network originating from the hyperlink structure of www.epa.gov~\cite{Batagelj_Mrvar_2006}. One can clearly see how the hubs in the classical algorithm tend to concentrate nearly all the importance. The insets show that the quantum algorithm is capable to lift the degeneracy of nodes in the the low part of the list~(see text in sect. \ref{sect_II})}
\end{figure}

Furthermore, we find that the hierarchy as predicted by the classical PageRank is not preserved. This is a property already found in~\cite{gdpmamd2011} for smaller networks. From our study we are able to clearly conclude that far from being an artefact of choosing small networks, this results to be a generic feature of the quantum PageRank algorithm.
We also found that the quantum PageRank is able to lift the degeneracy of the nodes that have a lower importance. This feature can be seen clearly in figure \ref{fig:MM_EPA_900_2300_Class_PR_vs_QPR} 
where we analyze a subgraph of the WWW obtained by exploring pages linking to www.epa.gov and available from Pajek~\cite{Batagelj_Mrvar_2006}.
%% 

%%%%%%%%%%%%%%%%%%%%%%%%%%%%%%%%%%%%%%%%%
%%%%%%%%%%%%%%%%%%%%%%%%%%%%%%%%%%%%%%%%%
\section{Quantum PageRank on Erd\H{o}s-R\'enyi Networks and Hierarchical Networks}
\label{sect_III}
%%%%%%%%%%%%%%%%%%%%%%%%%%%%%%%%%%%%%%%%%
%%%%%%%%%%%%%%%%%%%%%%%%%%%%%%%%%%%%%%%%%

\subsection{Erd\H{o}s-R\'enyi Networks}

In this section we will briefly introduce the  Erd\H{o}s-R\a'enyi class of random networks and analyze the performance of the quantum PageRank algorithm applied to them.

This class of graphs was introduced by Paul Erd\H{o}s and Alfred R\'enyi more than fifty years ago~\cite{Erdos_Renyi_1959,Erdos_Renyi_1960,Erdos_Renyi_1961}, and is of particular importance in the context of Graph Theory. 
There are different equivalent methods to describe this family.  To allow an easy computer implementation we use the following: a graph of order $N$  can be constructed by connecting $N$ vertices randomly by adding edges  with  a given probability,  which is independent from other edges.  We use directed versions of the graphs created with NetworkX~\cite{HaScSw08}.

The graphs falling into this class follow a Poissonian degree distribution, i.e.,  $P(k) \approx \langle k \rangle^k  \exp{(-\langle k \rangle )} /k!$ where $\langle k \rangle$ is the average degree. Thus, most nodes have a degree not far from the average and therefore the graphs do not display relevant hubs when applying the PageRank algorithm.

\begin{figure}
\includegraphics[keepaspectratio=true,width=.99\linewidth]{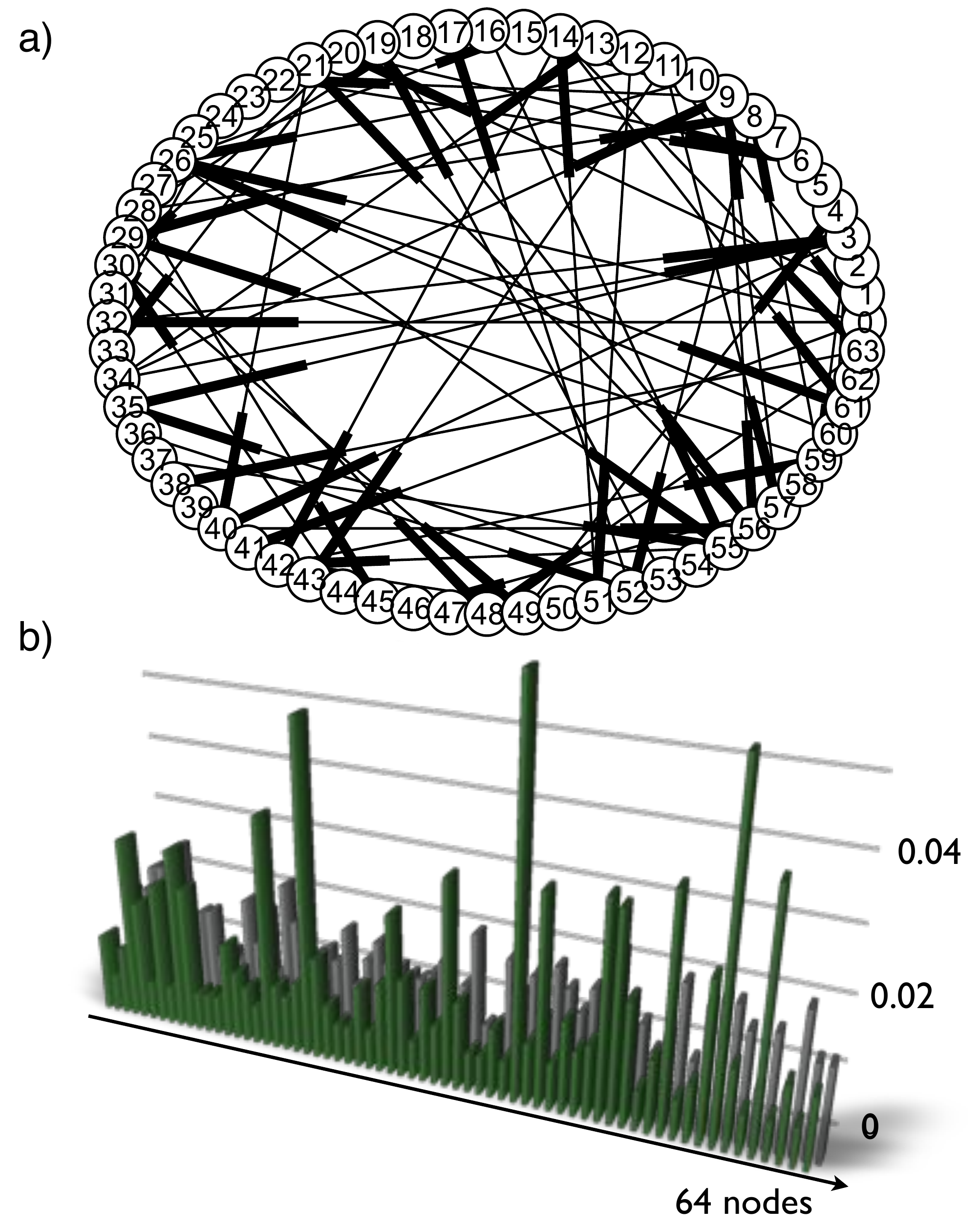}
\caption{
\label{fig:ERGraph64}
(Color online) Quantum and classical PageRank in random (Erd\"os-R\'enyi) networks. Subfigure a shows a prototypical example of a random network of $64$ nodes clearly indicating the absence of hubs in this class. b) A comparison of the importance as obtained from the quantum and classical PageRank applied to a $64$ nodes random graph. The importances calculated using the quantum PageRank algorithm display a change in hierarchy (see text in sect.~\ref{sect_III}). 
}
\end{figure}

We have performed a numerical study on graphs of the Erd\H{o}s-R\a'eny  type and found that the quantum PageRank algorithm displays a sharp change in hierarchy (see figure \ref{fig:ERGraph64} b). 
We can draw two fundamental conclusions: i/  The behavior of the quantum PageRank  algorithm is intrinsic and characteristic of the Erd\H{o}s-R\a'enyi  class of networks, in particular, the no-hub-behavior is reflected at the quantum level.
ii/ However, the quantum PageRank algorithm changes the ranking of the classical algorithm, in this sense outperforming the classical PageRank hierarchy. 
%% 
%%%%%%%%%%%%%%%%%%%%%%%
%%%%%%%%%%%%%%%%%%%%%%%
\subsection{Hierarchical Networks}
\label{subsect_Hierarchical}
%%%%%%%%%%%%%%%%%%%%%%%
%%%%%%%%%%%%%%%%%%%%%%%

Some relevant real-life networks which describe technological and biological systems, such as the WWW, some electronic circuits and protein or metabolic networks are usually scale-free but have also a modular structure ~\cite{Ravasz_2002,Song_2005}. That is, they are composed of modules that group different sets of nodes.
These modules can be distinguished  by the fact that nodes belonging to the same module are usually strongly connected. On the other side, modules are relatively weakly connected among them. Thus, even when the networks are scale-free, their hubs use to have a low clustering as they joint different modules. Several authors claim that a signature for a hierarchical network is that, other than the small-world scale-free characteristics,  the scaling of the clustering of the vertices of the graph with their degree follows $C_i \propto 1/k_i$~\cite{RaBa03,Barabasi-Oltvai-04}.

Hierarchical network models usually are constructed from  recursive rules. 
For example, we can start from a complete graph $K_n$  
and connect to a selected root node $n-1$ replicas of $K_n$.  
Next, $n-1$ replicas of the new whole structure are added to this root. 
At this step the  graph will have $n^3$ vertices. The process continues until we reach the desired graph order. 
There are many variations for these hierarchical networks, depending on the initial graph, the introduction of extra edges  among the different copies of the complete subgraphs, etc. However, once the starting graph is given, these networks do not have adjustable parameters and their main characteristics are fixed.

\begin{figure}
\includegraphics[keepaspectratio=true,width=.99\linewidth]{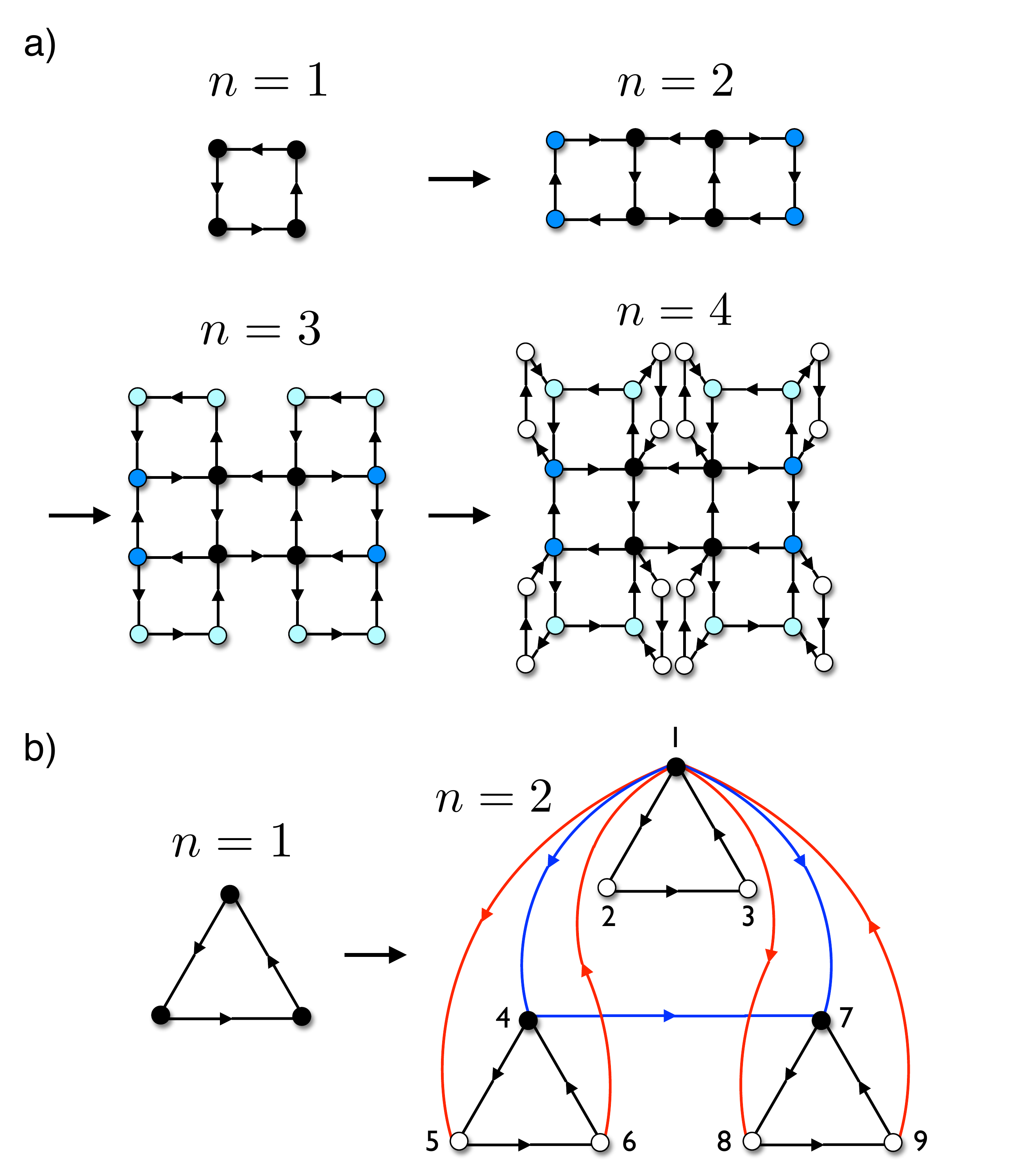}
\caption{
\label{fig:HierarchicalConstruction2}
(Color online) Construction of  hierarchical networks. a) The family of outerplanar directed hierarchical graphs. Note that the generation labeled by $n$ has $2^{n+1}$ nodes. We consider graphs of the generations with $n=4,5,6$.
b) The family of directed hierarchical graphs. In this case the generation labeled by $n$ has $3^{n}$ nodes. We consider graphs of the generations with $n=2,3,4$.
}
\end{figure}

In \cite{BaRaVi01}, Barab\'asi et al.  introduced a simple hierarchical family of 
networks and showed it had a small-world scale-free nature. The model was generalized in~\cite{RaBa03} and further studied in~\cite{No03}.   For our analysis we have designed a directed version based on these graphs, see figure~\ref{fig:HierarchicalConstruction2}b. In this case the starting point is a directed 3-cycle.

Another interesting family of hierarchical directed graphs has been obtained by giving directions to the edges of  the construction published in~\cite{Comellas_Miralles_PhysA,Comellas_Miralles_JPhys}, see figure~\ref{fig:HierarchicalConstruction2}a.
The graphs are in this case small-world, self-similar, unclustered and outerplanar (a planar graph is called outerplanar if it has an embedding where all vertices lie on the boundary of the exterior face). However, they are not scale-free, but follow an exponential distribution.
It has been shown that many algorithms which are NP-complete for general graphs perform polynomial in outerplanar graphs~\cite{BrLeSp99}.

\begin{figure}
\includegraphics[keepaspectratio=true,width=.95\linewidth]{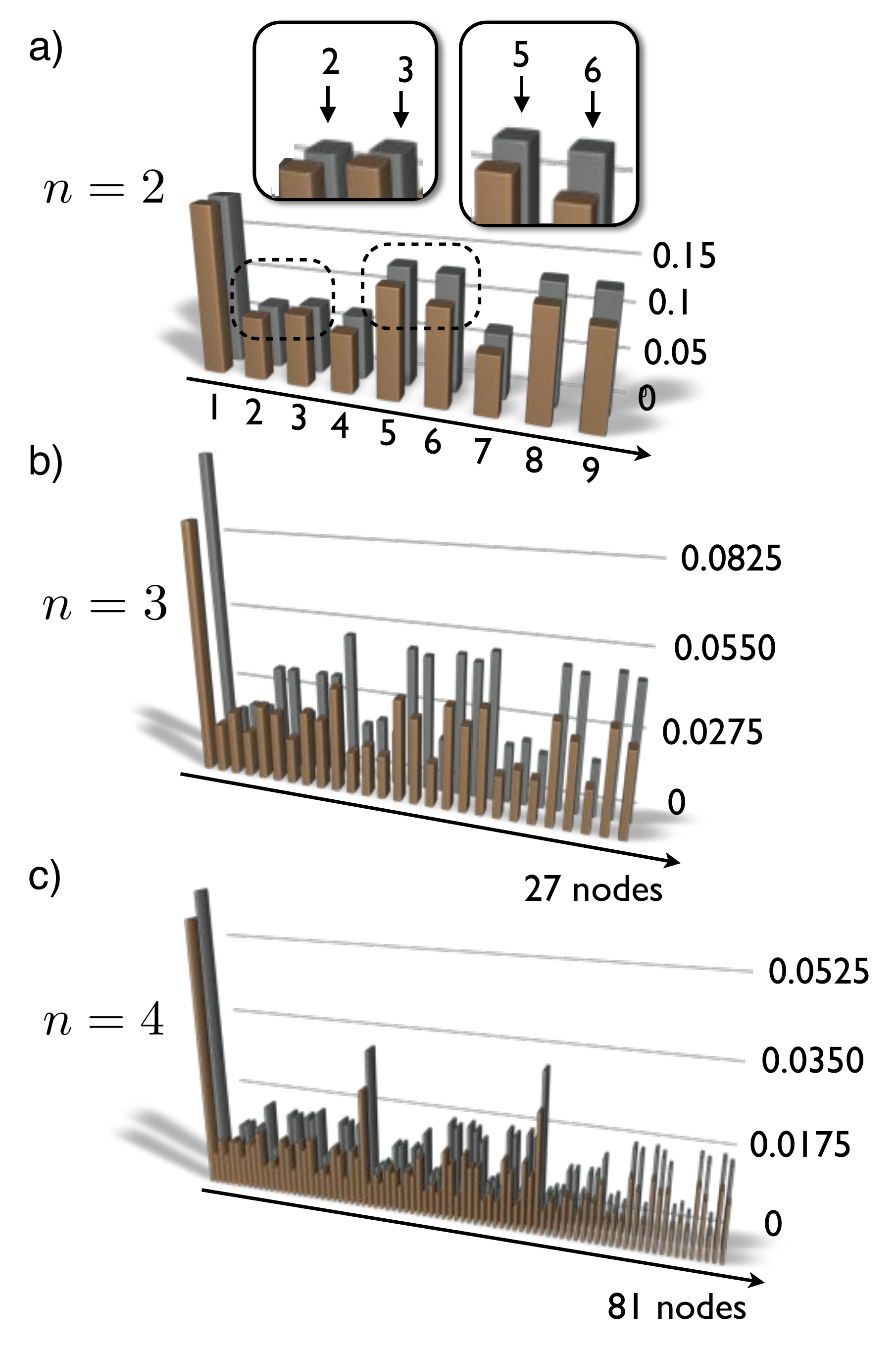}
\caption{
\label{fig:hierarchical_type1}
(Color online) Comparison of the quantum and classical PageRank for the family of hierarchical graphs described in the text (see section~\ref{subsect_Hierarchical} and figure~\ref{fig:HierarchicalConstruction2}b for the construction). We consider graphs with $n= 2,3,4$ (see subfigures a, b and c respectively). 
We find that the quantum PageRank preserves the hierarchy of the nodes but in addition it is able to highlight the connectivity structure of the nodes belonging to the same level. Indeed, in subfigure a, for example, the difference in importance between nodes $2$ and $3$ and $5$ and $6$ is amplified when calculated using the quantum algorithm.}
\end{figure}

\begin{figure}
\includegraphics[keepaspectratio=true,width=.9\linewidth]{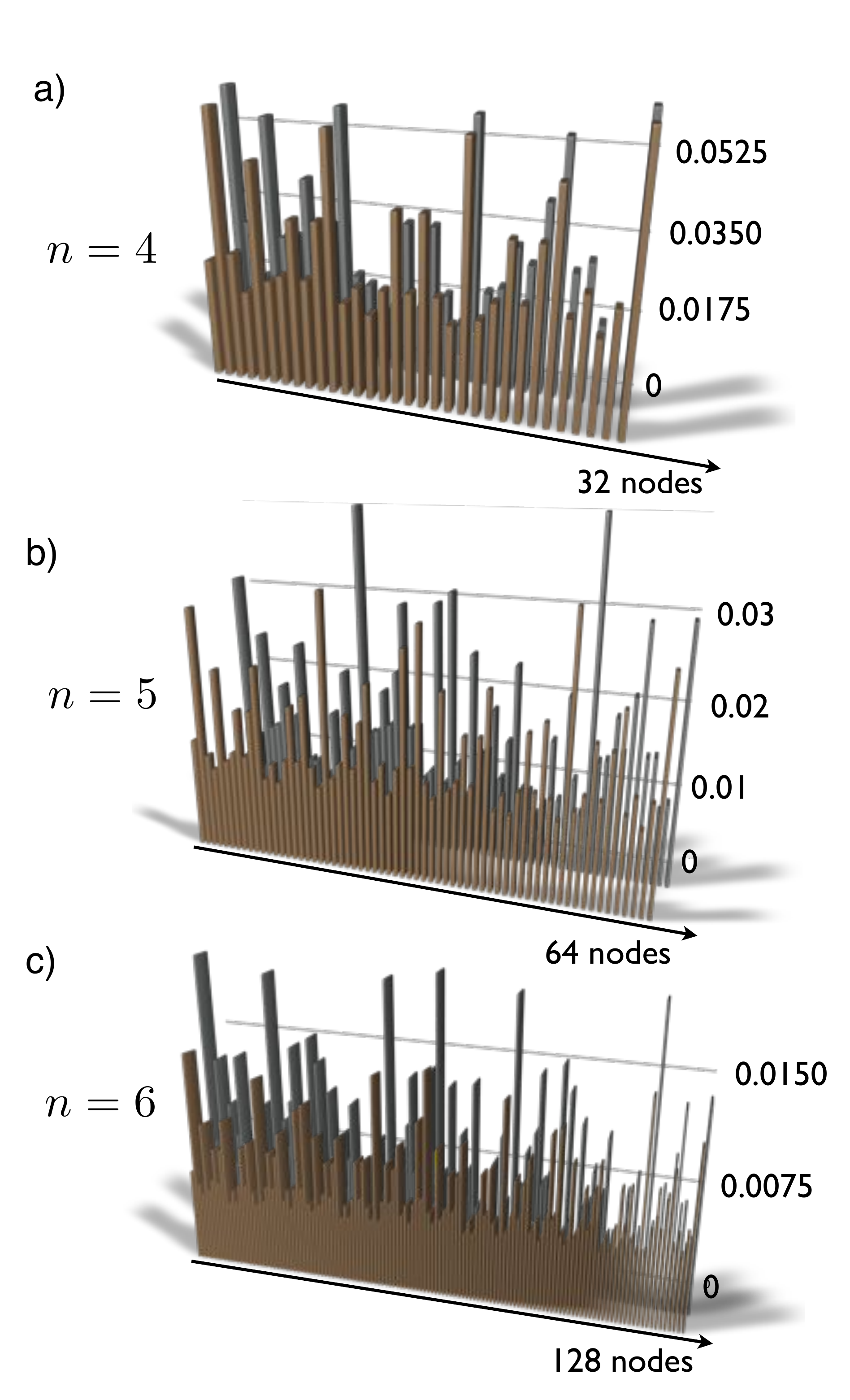}
\caption{
\label{fig:hierarchical_type2}
(Color online) Comparison of the quantum and classical PageRank for the family of hierarchical graph described in the text (see section~\ref{subsect_Hierarchical} and figure~\ref{fig:HierarchicalConstruction2}a for the construction). We considered graphs with $n= 4,5,6$ (see subfigures a,b and c respectively). 
Also in this case we find that the quantum PageRank preserves the hierarchy of the nodes highlights moreover the connectivity structure of the nodes belonging to the same level.}
\end{figure}

We have performed a numerical study on hierarchical networks using the quantum PageRank algorithm. We analyzed two families of graphs (see figure~\ref{fig:HierarchicalConstruction2} for the construction) and we find that the hierarchy (similarly to what was found in ref.~\cite{gdpmamd2011} for the binary tree) is preserved by the average PageRanks. Interestingly, though, the quantum PageRank is able to highlight the connectivity structure of the nodes that belong to the same level in the hierarchical construction (see figures~\ref{fig:hierarchical_type1} and~\ref{fig:hierarchical_type2}). We observe that  the difference in importance between nodes belonging to the same hierarchical level but with different local connectivity is amplified when calculated using the quantum PageRank algorithm.
%% 

%%%%%%%%%%%%%%%%%%%%%%%%%%%%%%%%%%%%%%%%%
%%%%%%%%%%%%%%%%%%%%%%%%%%%%%%%%%%%%%%%%%
\section{Localization-Delocalization Transition}
\label{sect_IV}
%%%%%%%%%%%%%%%%%%%%%%%%%%%%%%%%%%%%%%%%%
%%%%%%%%%%%%%%%%%%%%%%%%%%%%%%%%%%%%%%%%%

We have seen in the previous sections that the quantum PageRank algorithm is able to distinguish between networks of Erd\"os-R\'enyi and scale-free type. In particular, the quantum algorithm, as opposed to the classical one, is also able to highlight the secondary hubs in the case of scale-free networks. 
Furthermore, regarding the classical and quantum algorithms as walks on a network, a necessary condition that few nodes, the hubs, account for the majority of the importance is that the walker be localized, i.e. the number of nodes with a significant probability to find the walker is negligible with respect to the number of nodes in the network. 

\noindent In this section we will, therefore, study the localization properties of the walker on different networks, of Erd\"os-R\'enyi and scale-free type. 
We will briefly review the case of the walker in the classical PageRank algorithm and explore which phases does the quantum walker choose in the case of the quantum PageRank as a function of the structural properties of the underlying network.

\noindent In order to measure the localization phenomenon we will employ the Inverse Participation Ratio (IPR). This concept was introduced in the context of condensed matter and, more specifically, to study the Anderson localization phenomenon in disordered systems (see for example~\cite{evers2008}). It is particularly useful to study localization-delocalization transitions. 

\noindent The IPR $\xi_{cl}$, in the case of a classical walker, is defined as :
\begin{equation}
\xi_{cl} := \sum_{i =1 }^N \left[\mathrm{Pr}(X=i)\right]^{2r} ,
\label{eq:definition_IPR_classical}
\end{equation}
where $r > 0$ is an integer parameter which can be freely chosen and is fixed. 

\noindent Let us consider a classical walker on a network. We can have two extreme behaviors. The first one is that the walker is completely delocalized i.e. the probability of finding it on a site is uniform. Therefore, if we introduce a random variable $X$ whose realizations are the sites of the lattice, we can write $\mathrm{Pr}(X=i) = 1/N \, , \,  \forall i$. 
The other limiting case is that the walker is localized only on one site, i.e. the probability of finding the walker is a Kronecker delta, that is $\mathrm{Pr}(X=i) = \delta_{i j}$, if the walker is localized on say, site $j$. 
The IPR (eq.~\eqref{eq:definition_IPR_classical}) yields for the two limiting behaviors:
\begin{equation}
\xi_{cl} := 
\begin{cases} 
\; 1  & \mbox{if  the walker is localized}\\
\;N^{1-2r}   & \mbox{if the walker is delocalized. }
\end{cases}
\label{eq:IPR_behavior_classical}
\end{equation}
Thus the IPR, displaying respectively, a power law or a constant behavior as a function of the number of nodes, is a legitimate witness of the localization of the walker. 

\noindent We can rewrite $\xi_{cl} $ as: $ \xi_{cl} = N^{-\tau_{2r}}$. In order to study the localization-delocalization transition, it is useful to introduce the \empty{normalized} {\it anomalous dimension} $\Delta_{2r}$:
\begin{equation}
\tau_{2r} := (2r-1) + \Delta_{2r} ,
\label{eq:anomalous_dim_IPR_classical}
\end{equation}
which interpolates between the two phases if the system undergoes a transition from a localized regime (where $ \Delta_{2r} = 1-2r$) to a delocalized one (where $ \Delta_{2r} = 0 $). 

\noindent In \cite{shepelyansky} the localization-delocalization transition for a classical walker performing a random walk (with transition matrix given in \eqref{eq:_google_matrix_pagerank} ) was characterized studying its dependence on the damping parameter $\alpha$ where $0<\alpha<1$. This study is important to understand at a deeper level the classical PageRank algorithm. 
Indeed, as we have anticipated, in the case of a scale-free graph observing localization over a broad range of values for $\alpha$ is indeed a necessary condition for the algorithm, quantum or classical, to perform well the task of ranking the nodes.
The PageRank vector $I$, is given by:
\begin{equation}
G I =  I
\label{eq:classical_PR_eig_eq}
\end{equation}
and represents the stationary probability distribution of the walker on the network.

\noindent In ref.~\cite{shepelyansky} it was found that delocalization is absent for a very large range of values of $\alpha $ ranging from $0.4 $ to $1$.

\noindent This is quite natural because for the range of values of the damping factor $\alpha$ stated above, the second term, corresponding to random hopping in the Google matrix (see eq.~\eqref{eq:_google_matrix_pagerank}) is suppressed. This yields the localization effect. On the other side, for $\alpha$ close to $0$, it was found that the walker is delocalized over the network. This is understood easily, indeed for $\alpha $ close to  $0$ only the second term in the Google matrix survives and the walker is only subject to random uniform hopping between any pair of nodes. 
This leads to a trivial phenomenon of delocalization of the walker over the whole network.

Let us now focus on the localization phenomenon in the case of the quantum walk on a network according to the quantum PageRank protocol. In order to carry out the analysis we need to generalize the definition of the IPR given above reinterpreting the notion of probability of finding a walker on a node when dealing with the quantum PageRank.
In doing so, we will choose as a guiding principle the interpretation of the average quantum PageRank of a node (see eq.  \eqref{eq:average_QPR_def}) as the probability of finding the quantum walker on a particular node.
Therefore, we will employ the definition given in eq.~\eqref{eq:average_QPR_def} that we will rewrite as:
\begin{equation}
\langle I_q (P_i)\rangle = \left\langle \mathrm{Tr}_1  \mathrm{Tr}_2 \left( U^{2t} \rho(0) U^{\dagger \,2t} M^{(2)}_i \right)\right\rangle_t \, ,
\label{eq:recap_I_q_mean_in_time}
\end{equation}
where $M^{(2)}_i$ is the (strong) measurement operator on the second space (indexing the nodes where the edges point to, see subsect.~\ref{subsect_intro}), i.e. $M^{(2)}_i = |i\rangle_2 \langle i | $. 

\noindent We are now in a position to define the IPR $\xi$ in the case of a quantum walk: 
\begin{equation}
\xi := \sum_{i =1 }^N  \langle I_q (P_i)\rangle^{2r}\, .
\label{eq:definition_IPR_quantum}
\end{equation}

\noindent Also in the case of a quantum walk we have $\xi = N^{-\tau_{2r}} $ and it is evident that one can extract the localization phase of a walker from the scaling exponent of the IPR as a function of the number of nodes $N$. Indeed, from equation:
\begin{equation}
\log \xi \sim \left( 1-2r  -\Delta_{2r} \right)\log N
\label{eq:log_xi_vs_logN_quantum}
\end{equation}

\noindent it is clear that the witness of the localization lies in the slope of the graph of the aforementioned log-log plot.

We consider two kinds of networks, of the scale-free and of the Erd\"os-R\'enyi type. In order to study the localization phenomenon we generated networks with different numbers of nodes belonging to the two aforementioned classes. We then calculate the IPR (in the following we will fix the parameter $r = 1$) in order to understand whether for $\alpha =0.85 $ the quantum walker was localized or delocalized on the network. 

\begin{figure*}
	\includegraphics[width=\textwidth]{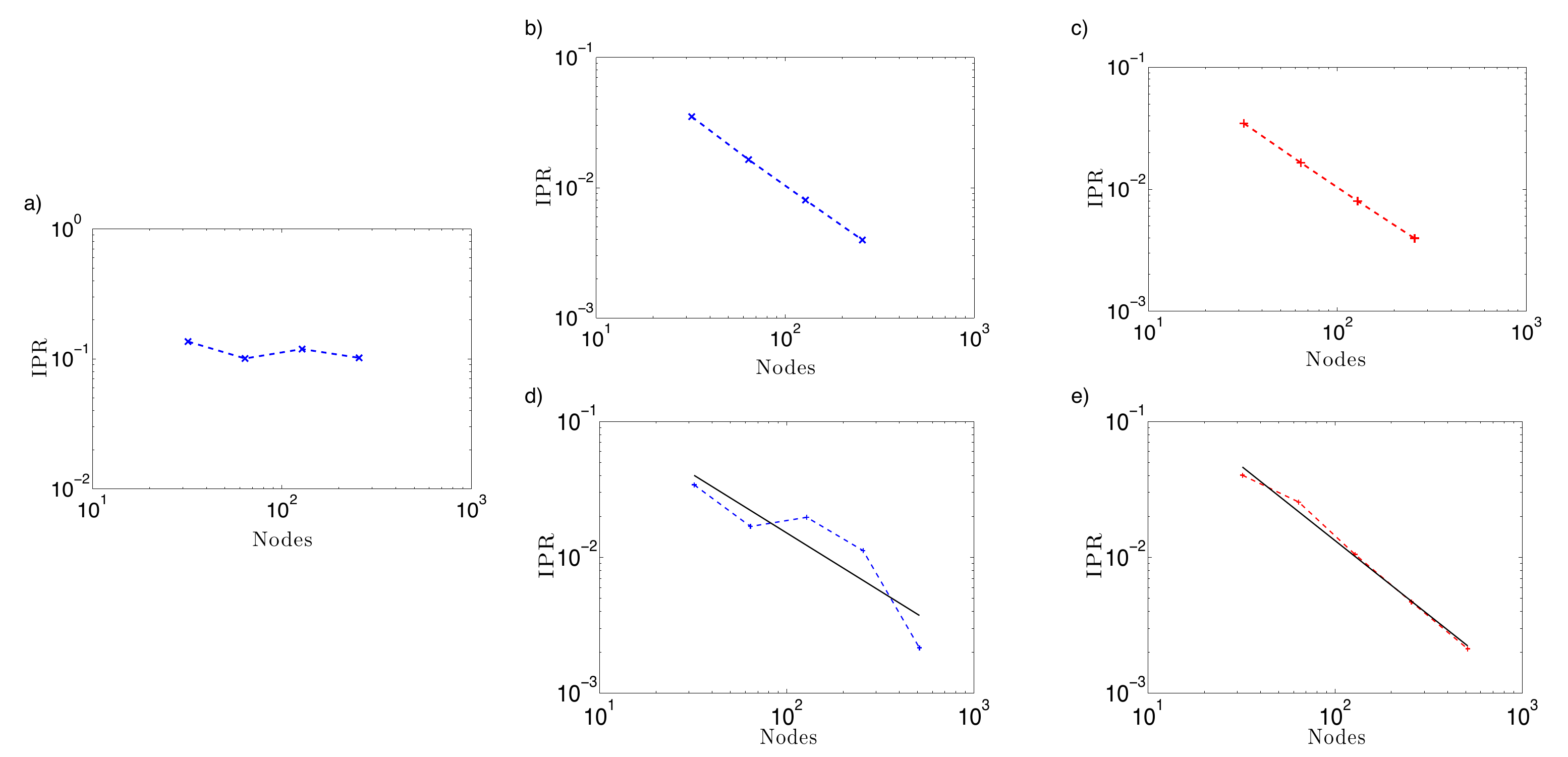}
	    \caption{(Color online) The IPR (for $r=1$) for networks of different classes, using the classical and the quantum walk, plotted versus the number of nodes in a log-log scale. a) The IPR using the quantum walk in the case of a scale-free graph and b) an Erd\"os-R\'enyi graph. c) The IPR using the classical walk in the case of an Erd\"os-R\'enyi graph. The value of $\alpha$ is set to $0.85$ in all cases.  In order to obtain the data we created using networkx 4 networks in the same class and with the same parameters having $32$, $64$,$128$ and $256$ nodes. We then calculated the IPR. In order to infer the phase of the walker we plotted $\log \xi $ vs. $\log N $. A constant behavior signals that the walker is localized whether a monotonically decreasing behavior signal delocalization over the network. We can infer that the quantum walker is in a localized phase in the case of the scale-free network (see (a)). This behavior is in contrast with what is displayed in the case of the Erd\"os-R\'enyi graph where delocalization is found in both the quantum and classical case (see subfigures (b) and (c)). This result holds true also when one considers graphs with a significantly lower link to node ratio. In this case we have superimposed two lines that result from a linear fit. These have equation $\log \xi = a_q \log N + b_q $ with $ a_q = -0.8565$ and $b_q = -0.2482$ for the quantum PageRank (see subfigures (d)). In the case of the classical PageRank  the equation is $\log \xi = a_{cl} \log N + b_{cl} $ with $ a_{cl} = -1.0932$ and $b_{cl} = 0.7125$ (see subfigures (e)). These results are consistent with delocalization (confront text in sect.~\ref{sect_IV}).}
 \label{fig:IPR_ER_SF_cl_q}
\end{figure*}

\noindent We find that the IPR in the case of the class of scale-free networks does not vary appreciably (see fig.~\ref{fig:IPR_ER_SF_cl_q}) signaling localization  of the walker on the graph. Notice however that also in ref.~\cite{shepelyansky} for these values of $\alpha$ a similar behavior was found.

\noindent We have analyzed also the graphs in the Erd\"os-R\'enyi class performing the same steps as above. Our study shows that, albeit this being random graphs and that for this value of the damping parameter $\alpha$ the walk is strongly influenced by the topology of the network, the quantum walker is delocalized in this case. Indeed it can be seen from figure~\ref{fig:IPR_ER_SF_cl_q} that the behavior of the logarithm of the IPR is  linear in the logarithm of the number of nodes. This is a clear witness of the delocalization phenomenon.

\noindent This behavior is remarkable for two reasons. The first reason is that in the classical random walk case localization of the walker was found~\cite{shepelyansky}. The second reason is that albeit randomness being expected to give rise to localization, in the case of graphs of Erd\"os-R\'enyi type, our study shows that the opposite is true. 

%%
%\begin{twocolumn}
%\end{twocolumn}
%%

\noindent We conclude that the scale-free graph seems to favor a localization phase in both walks, random and quantum, respectively for the classical and quantumPageRank algorithm. This result in the latter case was obtained with a value of $\alpha = 0.85$. This is consistent with a good ranking of nodes in a network. Indeed, in order to unveil the main hubs the random or quantum walk must be able to localize the walkers on few important nodes. 
Interestingly, instead, the Erd\"os R\'enyi graphs seem to prefer a delocalized phase albeit the networks being grown randomly. Both for the classical and the quantum PageRank one finds delocalization. This can be correlated with the absence of a small number of main hubs in this class of networks. 

\noindent As for the classical PageRank the localization-delocalization transition was characterized as a function of the damping parameter, it is important to understand how the quantum PageRank depends on the value of $\alpha$. We will study this dependence and characterize it in the next section.
%%%%%%%%%%%%%%%%%%%%%%%%%%%%%%%%%%%%%%%%%
%%%%%%%%%%%%%%%%%%%%%%%%%%%%%%%%%%%%%%%%%
\section{Stability of the Quantum Google Algorithm with respect to the Noise Parameter}
\label{sect_V}
%%%%%%%%%%%%%%%%%%%%%%%%%%%%%%%%%%%%%%%%%
%%%%%%%%%%%%%%%%%%%%%%%%%%%%%%%%%%%%%%%%%

In the previous section we analyzed the localization properties of the quantum PageRank algorithm. We studied this phenomenon having fixed the damping parameter using the value $\alpha = 0.85$. The natural question is to study how the quantum PageRank varies with respect to the variation of $\alpha$.

\noindent The stability of the quantum PageRank is an important issue to consider also because the damping parameter is \emph{arbitrarily} tuned to a specific value. Indeed there is no \textit{a priori} argument to fix the value of $\alpha $ and the value $0.85$ was originally chosen in the classical PageRank protocol to mimic  the behavior of a surfer (or a random walker) that randomly hops to any other page once every seven times. Only \textit{a posteriori} it turned out that this is indeed a sensible choice given that the network is small-world, being in fact a crucial ingredient for the PageRank algorithm to yield reasonable ranking results. In view of the ad-hoc choice of the precise of the parameter $\alpha$, it is a very desirable property that the output of the algorithm be stable, i.e. the ranking vary slowly with respect to the variation of the damping parameter. A question that was addressed in the computer science community has been to quantify how susceptible to changes in this parameter the classical PageRank algorithm is. It was found that the effect of this parameter on ranking is large and that two rankings obtained by running the algorithm using different values of this parameter can be very different~\cite{shepelyansky}.

\noindent To tackle this problem in the quantum case we will make use of two quantities. The first one is related to the \textit{ quantum fidelity} (see e.g.~\cite{rmp,NC}) that provides a way to measure the \textit{distance} between two quantum states. The second quantity that we will use is the \textit{classical fidelity}. It is employed for the same task when dealing with probability distributions.

In ref.\cite{shepelyansky} the stability of the classical PageRank was studied as a function of the damping parameter. 
\noindent Since the PageRank vectors are classical probability distributions one can measure the distance between two PageRank vectors, calculated using different values of the damping parameter, with the classical fidelity. The latter can be written as:
\begin{equation}
f(\alpha , \, \alpha ^\prime) = \sum_j \sqrt{I(P_j , \alpha) I(P_j , \alpha^\prime)}
\label{eq:fidel_Shep_def}
\end{equation}
It was found that there is a plateau around $(\alpha , \, \alpha ^\prime) \approx (0.5,\, 0.5)$ and that the fidelity $f(\alpha , \, \alpha ^\prime = 0.85) $ is rather flat around $\alpha \approx 0.85 $ thus implying that the classical PageRank is rather robust against perturbations (see ref.~\cite{shepelyansky}).

\noindent The \emph{quantum fidelity} is a quantity that measures the distance of two quantum states:
\begin{equation}
F(\sigma , \rho ) = \sqrt{\rho^{1/2} \, \sigma\,\rho^{1/2} }
\label{eq:fidel_quant_def}
\end{equation}
which in the case of commuting density matrices reduces to the classical fidelity.

\noindent Another valid measure of the distance of two quantum states is the {\it trace distance}:
\begin{equation}
D(\sigma , \rho ) = \frac{1}{2} \textrm{tr} | \rho - \sigma |
\label{eq:trace_distance_quant_def}
\end{equation}
where $|\tau | $ denotes the square root of the (positive) operator $\tau^\dagger \tau$.

\noindent The fidelity and the trace distance turn out to be equivalent measures of { \it distance}. Indeed if the fidelity of quantum states is near to one then their trace distance is close to zero and viceversa \footnote{This can be seen from a general formula that relates the two measures of distance (see e.g. \cite{NC} 
chap. 9): $1- F(\sigma , \rho )  \le D(\sigma , \rho ) \le \sqrt{ 1 - F(\sigma , \rho )^2} $}
%\cite{footnote_fidelity}. 
%
Therefore, either can be used for our purpose of measuring the stability of the quantum PageRanks and the choice is a mere matter of convenience. To tackle the problem of stability in the case of the quantum PageRank we will use the {\it trace distance} and the classical fidelity.

\noindent Let us rewrite explicitly the definition of the averaged quantum PageRank $\langle I_q (P_i,\alpha) \rangle$ adding the dependence on $\alpha $ (that enters in the initial state and in the evolution operator of the walk) as:
\begin{equation}
\langle I_q (P_i, \alpha) \rangle= \left\langle \mathrm{Tr}_1 \left( \mathrm{Tr}_2 \left[  \rho^{12}_\alpha(2t) M^{(2)}_i \right]\right)\right\rangle_t
\label{eq:recap_I_q_alpha_dep}
\end{equation}
where $ \rho^{12}_\alpha(2t) =   U_\alpha^{2t} \rho^{12}_\alpha(0) U_\alpha^{\dagger \,2t}$ and for bookkeeping purposes it has been made explicit to which spaces the density matrix refers to. 

Let us discuss how to apply the concept of trace distance in our case. We might measure the {\it instantaneous} distance $D(\rho_\alpha (2t) ,\, \rho_{\alpha^\prime} (2t))$, thus measuring the distance of the quantum states. 
However, it is more significant to take the time-average of the distance between the partial traces of the states, because we are interested in the quantum PageRank as an observable, rather than the state itself.
Not being interested in the distance of \emph{states}, we can make use of a less refined measure of distance (that is, one that can appreciate less the difference of two states). This can be written in the following form  (see the appendix for the derivation):
\begin{align}
  \left\langle D \left( \mathrm{Tr}_1 \rho^{12}_\alpha(2t) , \mathrm{Tr}_1 \rho^{12}_{\alpha^\prime}(2t) 
 	\right)  \right\rangle_t  
	= \nonumber \\ =
	 \max_i  \left|
	  \langle I_q(P_i,t,\alpha) \rangle -\langle I_q(P_i,t,\alpha^\prime) \rangle 
	  \right|
\label{eq:trace_dist_stability_Av_QPR}
\end{align}

\noindent To summarize, we will study the stability making use of the classical fidelity and the quantity in~\eqref{eq:trace_dist_stability_Av_QPR}, that is a simpler measure descending from the trace distance (which turns  out to be equivalent to the quantum fidelity).
\begin{figure*}	
	\centering
	\includegraphics[keepaspectratio=true,width=.42\linewidth]{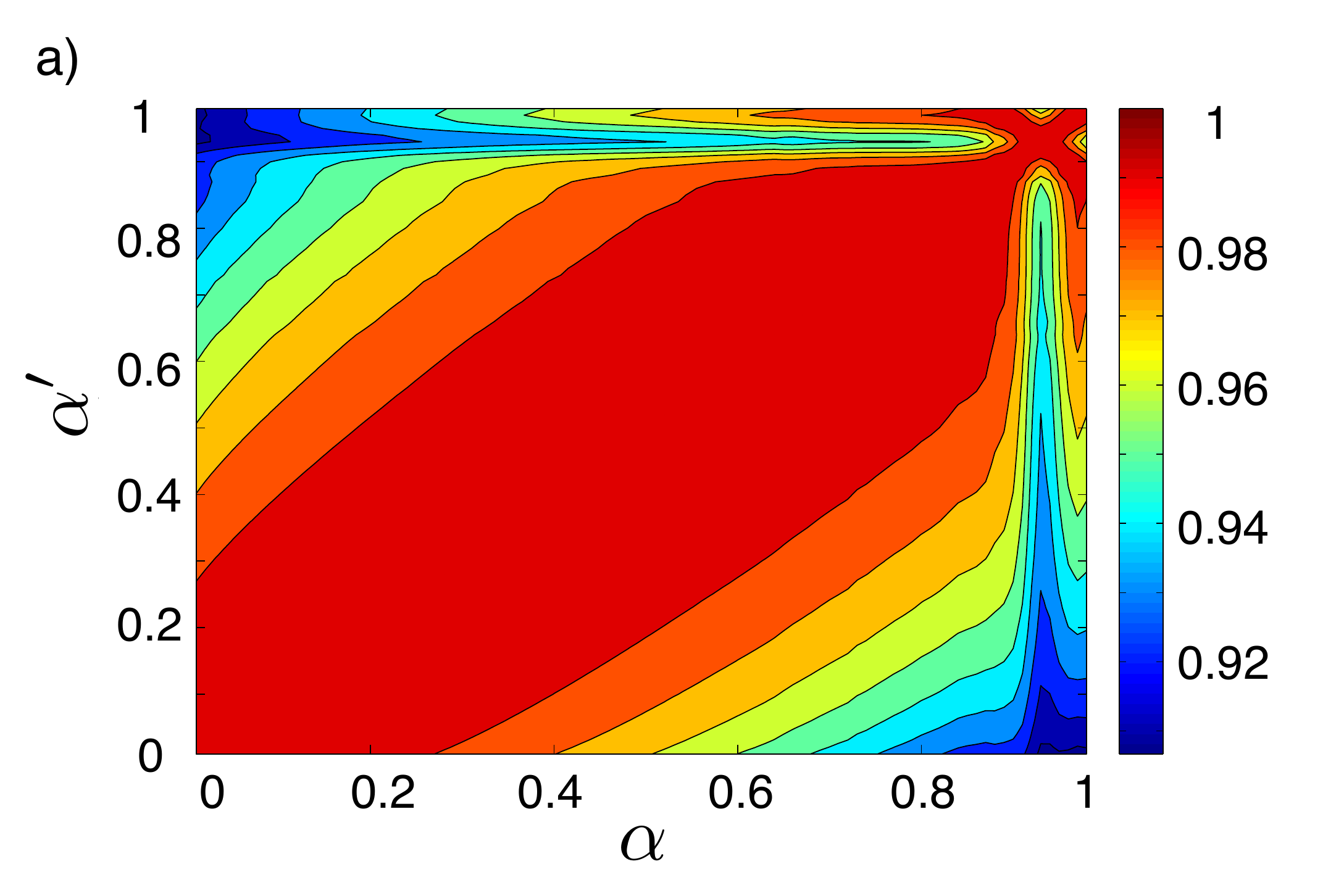}
        	 \includegraphics[width=0.40\textwidth]{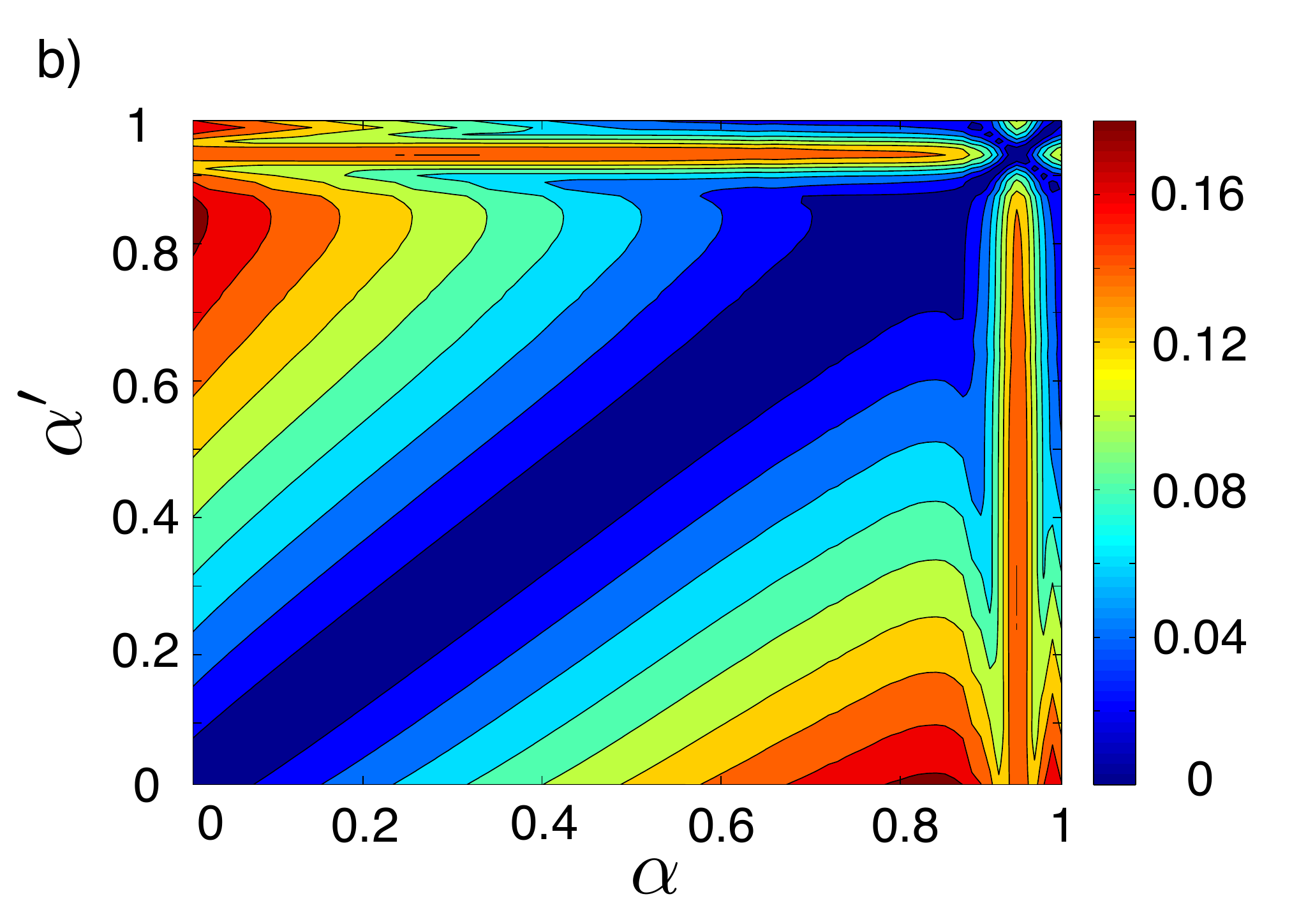}
%	}
\caption{(Color online) Measures of distance of quantum PageRanks obtained with different values of the damping parameter. 
The network analyzed is a scale free network with 128 nodes (generated with NetworkX). The damping parameter varies ranging from $0.01$ to $0.98$.
a) The fidelity obtained by applying the classical fidelity (see eq.~\eqref{eq:fidel_Shep_def}).  One can see that there is a plateau for values of $\alpha $ around $0.8$. 
b) The measure of distance obtained from the trace distance (see~\eqref{eq:trace_dist_stability_Av_QPR}). We obtain a similar result: a plateau for values of $\alpha $ around $0.8$ is clearly visible.}
\label{fig:Cl_fidelity_trace_dist_stability}
\end{figure*}
%%
%\begin{twocolumn}
\begin{figure*}	
	\centering
           \includegraphics[width=0.8\textwidth]{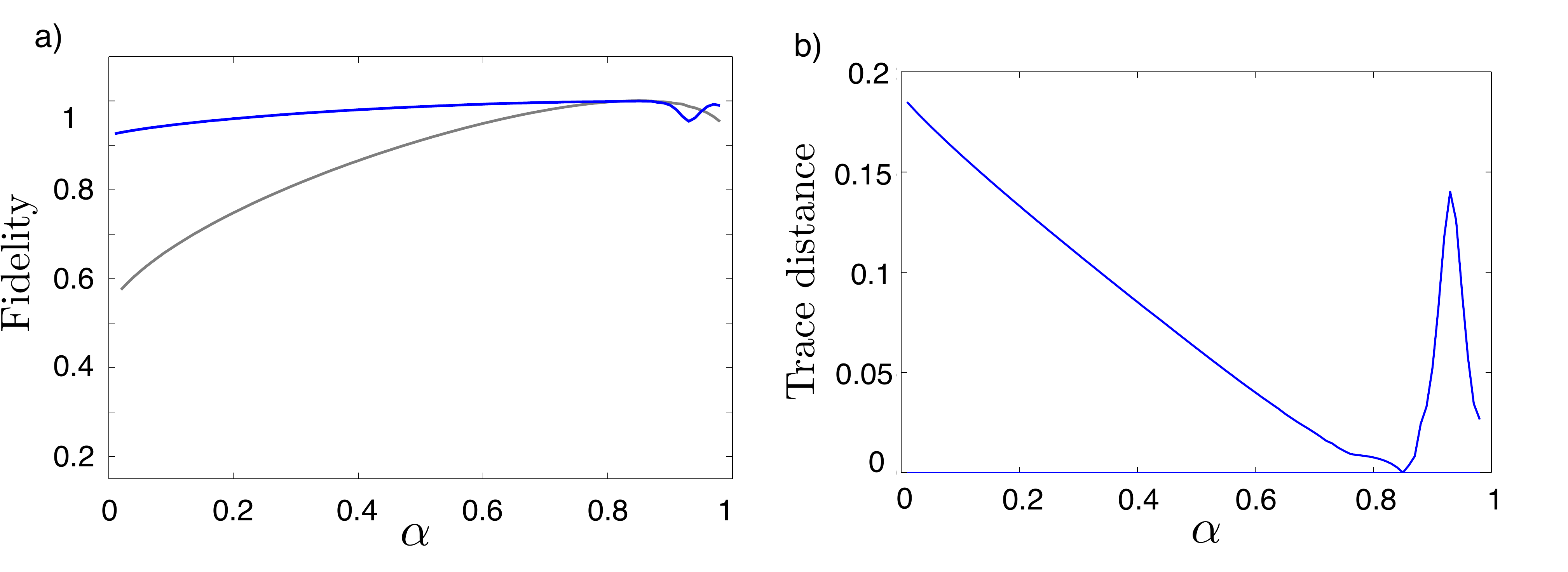}
	\caption{(Color online) Analysis of the stability with respect to the variation of the damping parameter from the value $\alpha = 0.85$. The network is a scale free graph with $128$ nodes generated using networkx for the quantum algorithm (colored lines) and of 256 nodes for the classical one.
a) In color the classical fidelity (see~\eqref{eq:fidel_Shep_def}) between the quantum PageRank calculated using $\alpha = 0.85$ and the one calculated using values in the range from $0.01$ to $0.98$. In grey the one for the classical algorithm. We notice that the quantum PageRank is more robust with respect to the variation of the damping parameter.
b)The trace distance between the quantum PageRank calculated using $\alpha = 0.85$ and the one calculated using values in the range from $0.01$ to $0.98$ (using~\eqref{eq:trace_dist_stability_Av_QPR}). Plot of the trace distance to compare the quantum PageRank obtained with $\alpha = 0.85$ to the one obtained using other values of the damping parameter.}
\label{fig:Fidelity_Both_Stability_CPR_vs_QPR_around_0_85_SF_128_nodes}
\end{figure*}
%\end{twocolumn}
%%
We perform an analysis of how the quantum PageRank varies with respect to the rankings when the value of $\alpha$ goes from $0.01 $ to $0.98$. 
We analyze a scale free with 128 nodes that was generated with networkx, a python module.  The results show clearly that the quantum PageRanks vary very little when the quantum walk underlying the quantum PageRank has a different damping parameter $\alpha$. Indeed it can be seen from figure~\ref{fig:Cl_fidelity_trace_dist_stability} that the minimum fidelity between two values of $\alpha$ and $\alpha^\prime$ does not go below the value of $0.91$. One should compare with the analysis of the classical PageRank~\cite{shepelyansky}, where the fidelity between different values of  $\alpha$ and $\alpha^\prime$ can be approximately $0$. It can thus be inferred that the ranking is rather robust when it is performed with the quantum PageRank.

\noindent We have also investigated the behavior around the value of the damping parameter $\alpha = 0.85$. From the classical fidelity between the quantum PageRanks at $\alpha = 0.85$ and at $\alpha $ ranging from $ 0.01$ to $ 0.98$ one can see that there is a plateau around the value of $\alpha = 0.85$ (see figure~\ref{fig:Cl_fidelity_trace_dist_stability}) extending especially for smaller values of $\alpha$. There is a dip for $\alpha = 0.95$ which is due to the fact that the ranking is very sensitive to changes in the damping parameter $\alpha$ when this approaches $1$. For this value only the second term in the Google matrix $G$, giving random hopping, survives.
	
\noindent The analysis was made more precise and the conclusions more cogent by considering also the measure of distance of rankings originating from the trace distance (see~\eqref{eq:trace_dist_stability_Av_QPR}). Also in this case the overall robustness of the ranking performed with the quantum PageRank is evident. One can see from figure~\ref{fig:Cl_fidelity_trace_dist_stability}  that the maximum value of this measure of distance is $0.18$ comparing any two values of $\alpha $ and $\alpha^\prime$ ranging in the aforementioned interval. It can be observed also that the region where this ranking is more robust is where $\alpha \approx 0.8$. Indeed from  figure~\ref{fig:Cl_fidelity_trace_dist_stability} the blue region is wider (and correspondingly the ranking more robust with respect to perturbation of  the value of $\alpha$).

\noindent We find also in this case that  figure~\ref{fig:Fidelity_Both_Stability_CPR_vs_QPR_around_0_85_SF_128_nodes} the trace distance is rather smooth for $\alpha = 0.8$. One can see that also in this case there is a curious peak for  for $\alpha \approx 0.95$ similarly to the previous case.

\begin{figure}
\includegraphics[keepaspectratio=true,width=.95\linewidth]{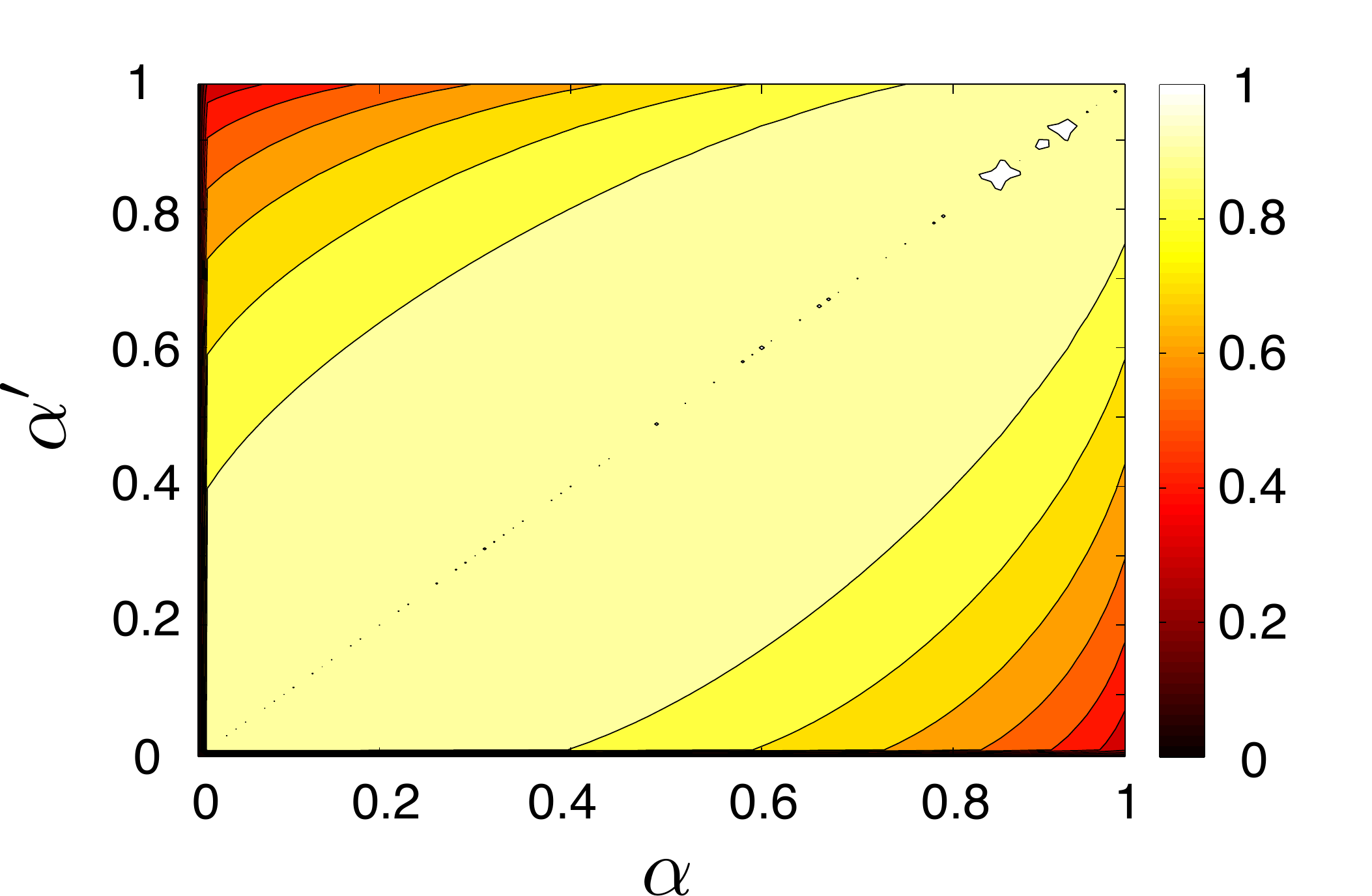}
\label{fig:Cl_Fidelity_Contourf_CLASSICAL_PR_Stability_SF_256_nodes}
\caption{(Color online) The fidelity obtained by applying the classical fidelity (see eq.~\eqref{eq:fidel_Shep_def}) in the case of the classical PageRank.
The network analyzed is a scale free network with 256 nodes (generated with NetworkX). The damping parameter varies ranging from $0.01$ to $0.98$. One can see that for extremal values of $\alpha$, $\alpha^{\prime}$ the value of the fidelity drops below $0.4$.}
\end{figure}

\noindent We conclude that the quantum PageRank as measured by the classical fidelity or by the trace distance seems to vary mildly with respect to the variation of the damping parameter $\alpha$. Indeed the minimum fidelity between any two distributions of importance arising from the quantum PageRank is $90\%$. The maximum of the distance obtained by using the trace distance between any two states (with different $\alpha$) is $0.18$.
This means that the quantum PageRank is very robust with respect to variation of the parameter that controls the fraction of random hopping. 
It is much more robust than in the classical case. From our analysis of the stability of the classical PageRank we find that for extremal values of $\alpha$, $\alpha^{\prime}$ the value of the fidelity is less than $0.4$ (see fig.~\ref{fig:Cl_Fidelity_Contourf_CLASSICAL_PR_Stability_SF_256_nodes}). In ref.~\cite{shepelyansky} the minimum fidelity was found to be very close to $0\%$ between PageRanks' rankings corresponding to extremal values of the damping parameter $\alpha$.

%%%%%%%%%%%%%%%%%%%%%%%%%%%%%%%%%%%%%%%%%
%%%%%%%%%%%%%%%%%%%%%%%%%%%%%%%%%%%%%%%%%
\section{Power Law Behavior for Quantum PageRank}
\label{sect_VI}
%%%%%%%%%%%%%%%%%%%%%%%%%%%%%%%%%%%%%%%%%
%%%%%%%%%%%%%%%%%%%%%%%%%%%%%%%%%%%%%%%%%x

In section~\ref{sect_II} we have found that the quantum PageRank is able to highlight the structure of secondary hubs. 

In addition to this finding we have shown in section~\ref{sect_IV}  that this strength of the quantum algorithm is associated with the fact that the quantum walker on networks of scale-free type is in a localized phase, i.e. the nodes with a significant average quantum PageRank are a negligible fraction of the nodes in the network.
On the other side, for the classical algorithm, it has been shown~\cite{donato2004,Pandurangan2005} that for real networks the nodes' classical PageRanks $I_j$, sorted in descending order, behave following a power law. That is, the classical PageRanks decrease approximately like $I_j \sim j^{-\beta}$, where $\beta \approx 0.9$.
This is a clear witness of the fact that the algorithm is able to identify the hubs. Furthermore, the scaling exponent $\beta$ measures the relative importance given to the hubs with respect to the other nodes.

\noindent A similar study is therefore important in the case of the quantum PageRank and will be carried out in this section. Indeed, it is desirable that the quantum PageRank display a power law scaling behavior. Indeed, such a behavior is distinctive of the fact that the algorithm be able to uncover the scale free nature of the network.

\noindent In order to point out the scaling behavior of the quantum PageRanks of the nodes it is clear from the conjectured form
\be
\langle I_q (P_i)\rangle \sim i^{-\beta_q}
\label{eq:scaling_QPR}
\ee
that we can extract such behavior from the slope of the log-log plot of the quantum PageRanks versus the (sorted) index of the  $i$ as can be seen from $
\log  \langle I_q (P_i)\rangle \sim -\beta_q \log (i) \, . 
$

\noindent Here we perform the analysis on scale free networks of 256 nodes. We calculate the classical and quantum PageRanks of the nodes and after having sorted the nodes in descending order we analyze the log-log plot of the classical and quantum PageRanks versus the nodes' index.

\begin{figure*}	
	\centering
           \includegraphics[width=0.49\textwidth]{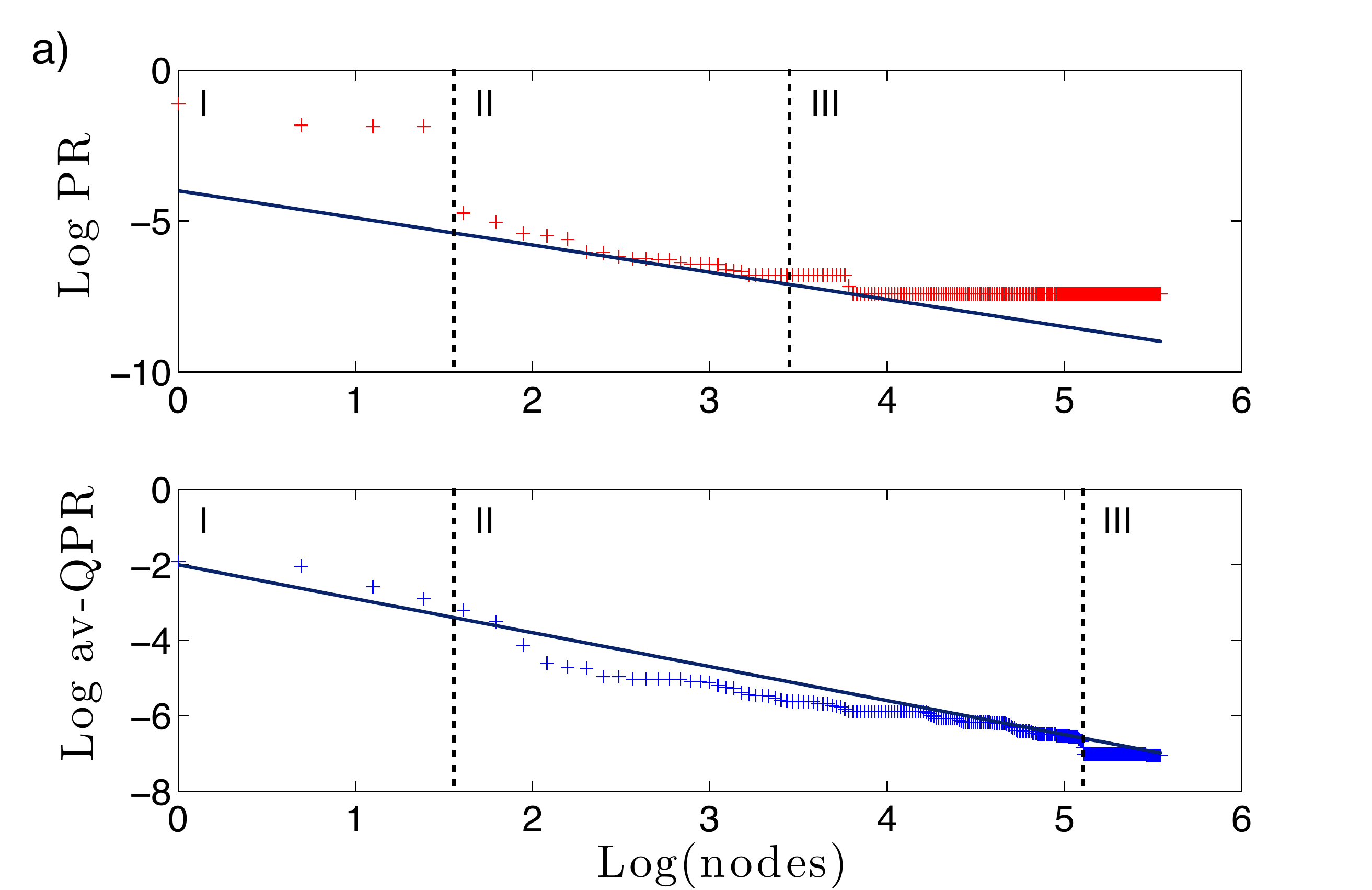}
           \includegraphics[width=0.49\textwidth]{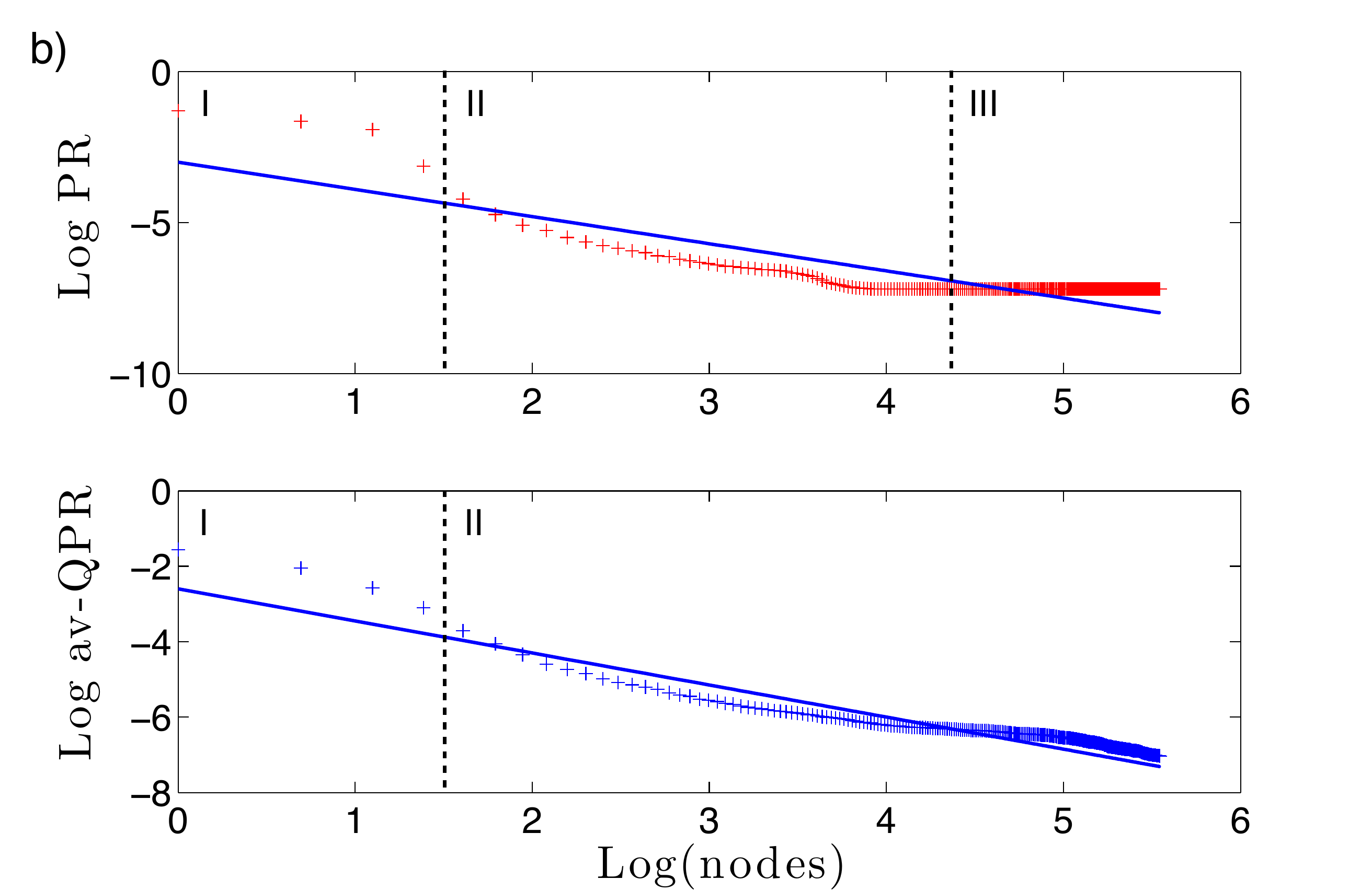}
%	}%
	\caption{(Color online) a) The plot of the logarithm of the PageRanks (upper part) and quantum PageRanks (lower part), (after being reordered, see text in section~\ref{sect_VI}) versus the logarithm of the node's label. As a guide to the eye we have superimposed two lines with slope equal to $-0.9$. 
One can clearly distinguish three zones (see text in section~\ref{sect_VI}). b) The plot of the logarithm of the means over the ensemble of graphs in a class of scale free networks of the PageRanks (upper part) and quantum PageRanks (lower part), (after being reordered, see text in section~\ref{sect_VI}) versus the logarithm of the node's label. As a guide to the eye we have superimposed two lines with slope equal to $-0.9$ in the classical case and $-0.85$ in the quantum case.
	}%
\label{fig:Beta_SF_256_nodes_Q_and_Cl20_and_mean}
\end{figure*}

\noindent Considering one instance of a graph in this class one can clearly see (cf. figure \ref{fig:Beta_SF_256_nodes_Q_and_Cl20_and_mean}) that both rankings obtained using the classical and the quantum PageRank display a power law behavior. The plot displays three areas, corresponding respectively to the hubs, to the intermediate part of the list and  the low part of the ranking (which in the classical case is degenerate in importance).

The fact that this behavior persists also in the case of the quantum PageRank is due to the effect, similar to what has been observed in~\cite{shepelyansky} for the classical case (and in section~\ref{sect_IV} for the quantum Pagerank that the walker is in a localized phase in the case of scale free networks). Consequently the most highly ranked nodes tend to concentrate nearly the totality of the importance. This can be clearly seen in figure~\ref{fig:Beta_SF_256_nodes_Q_and_Cl20_and_mean}, area $I$, where the hubs' importances lie above the line.

\noindent Furthermore, it is clear that the scaling coefficients are different in the quantum and classical case. We find that $\beta_q < \beta_{cl}$ the quantum PageRank has a smoother behavior, giving less relative importance to the nodes in the high part of the list. On the other hand it is also able to better rank in the low part of the list (where the classical PageRank gives highly degenerate values) lifting the degeneracy. (cf. the  area $III$ in figure~\ref{fig:Beta_SF_256_nodes_Q_and_Cl20_and_mean}).

The quantum  PageRank is therefore able to better distinguish the nodes' importances in the lowest part of the list. This is because the difference between the importance of the nodes in the higher and lower part of the list is lower. Furthermore, the power law behavior interpolates a greater portion of the data with respect to the classical case as can be noted by inspection (the area $II$ in figure~\ref{fig:Beta_SF_256_nodes_Q_and_Cl20_and_mean} extends much more in the case of the quantum algorithm).

\noindent To complete the analysis we consider an ensemble of scale free networks in order to display the ensemble's properties rather than the particular instance's. 
The ensemble consists of 29 scale free networks.  It is clear that the properties found in the instance of the graph in figure~\ref{fig:Beta_SF_256_nodes_Q_and_Cl20_and_mean} persist also when considering a mean property of the ensemble (see figure~\ref{fig:Beta_SF_256_nodes_Q_and_Cl20_and_mean}). We conclude that these properties are generic and not an artifact of considering only one instance of the class of scale-free networks.

\begin{figure}	
	\centering
          \includegraphics[width=0.49\textwidth]{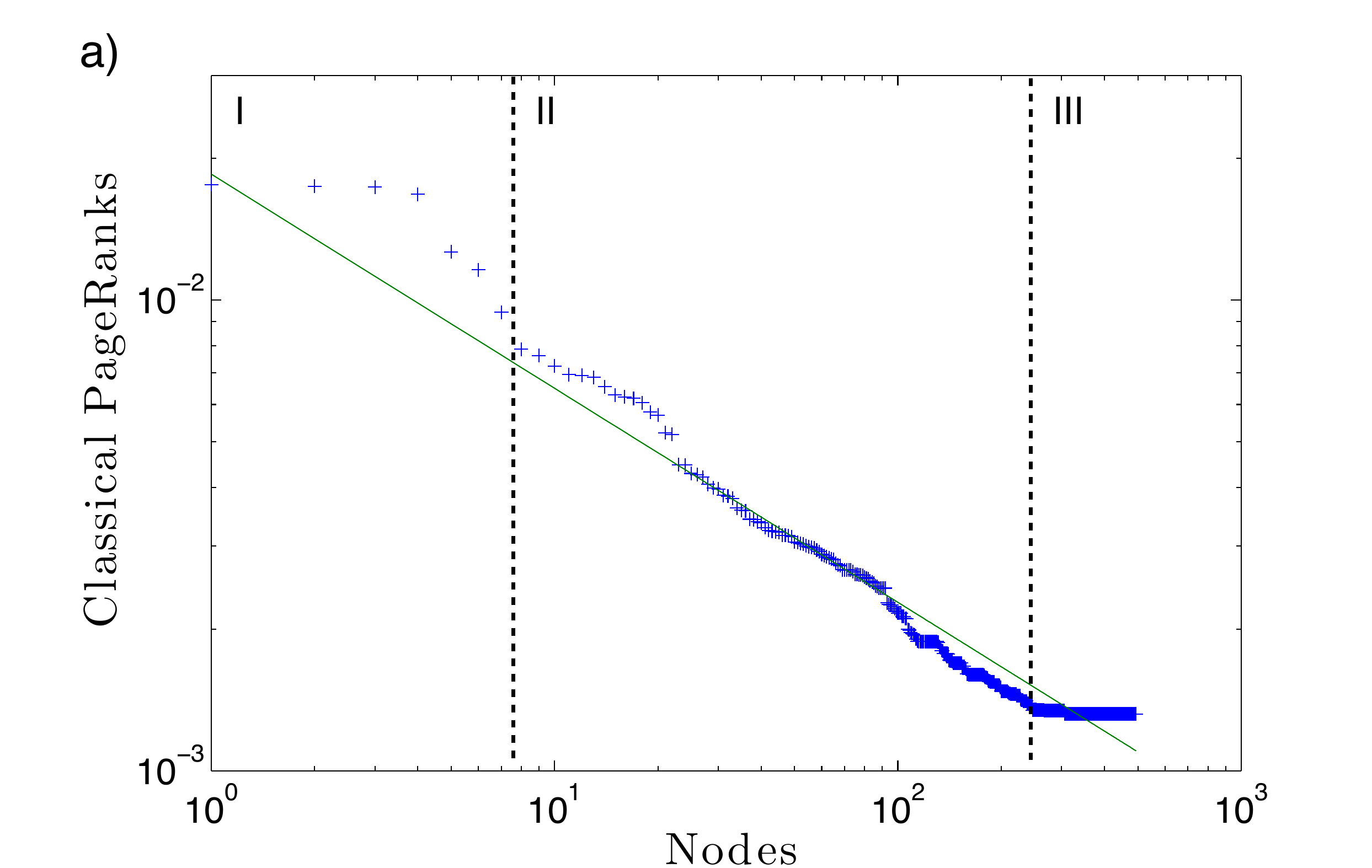}
          \includegraphics[width=0.49\textwidth]{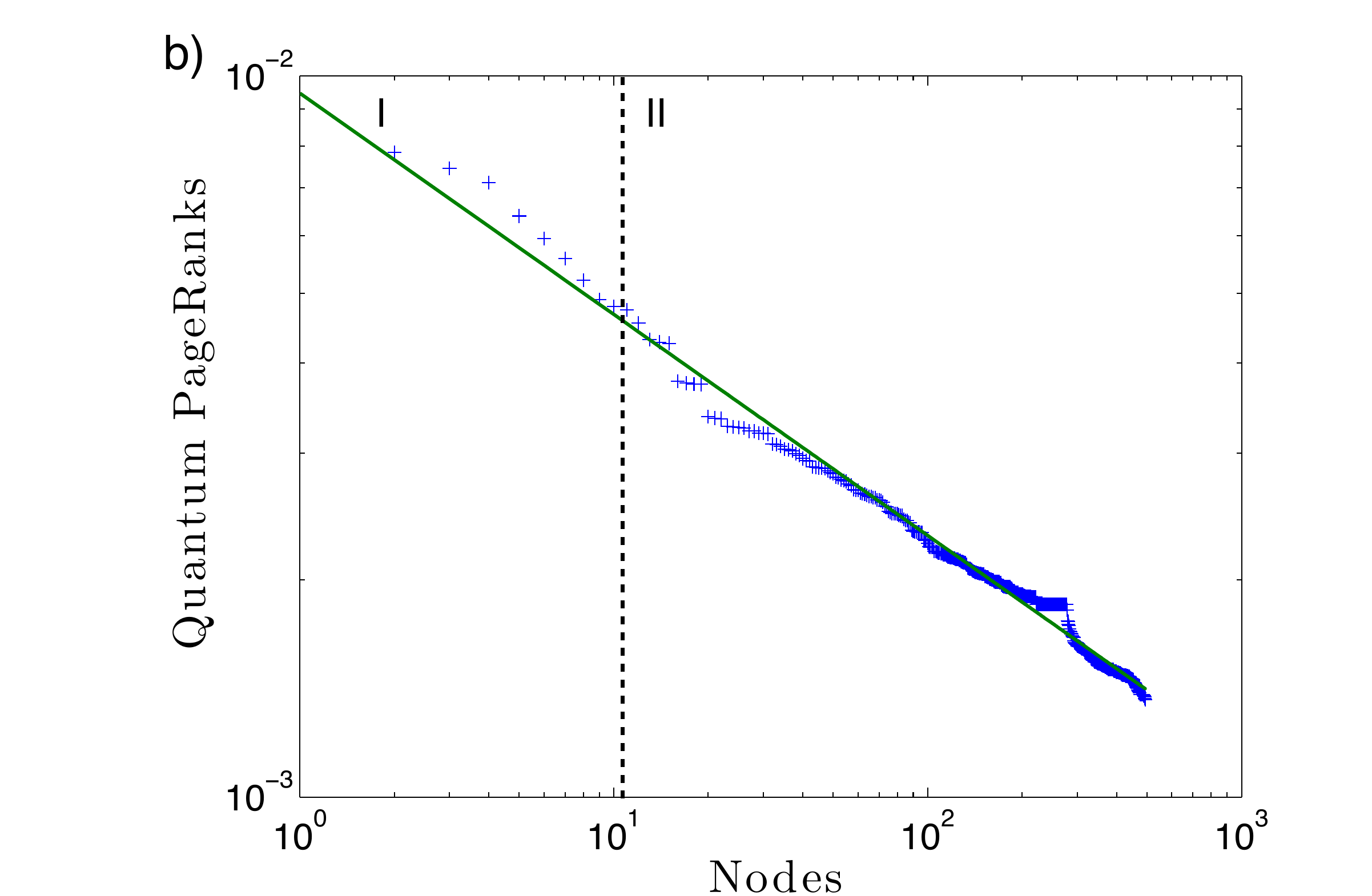}
%	}%
	\caption{(Color online)  Plot in log-log scale of the quantum and classical PageRanks of the nodes of a piece of a real network (from EPA). The quantum and classical PageRanks are displayed after having been sorted in descending order, see text in section~\ref{sect_VI}) versus the logarithm of the node's label. 	
a)  The classical PageRanks. As a guide to the eye we have superimposed the fitted lines (in log-log scale) with equation $I(i) = c_{cl} i^{-\beta}$ where $ \beta = 0.4545$ and $c_{cl} = 0.0185$.
b) The average quantum PageRanks. In this case we have superimposed the fitted line (in log-log scale) with equation $I(i) = c_{q} i^{-\beta}$ where $ \beta = 0.3066$ and $c_{cl} = 0.0095$.}
\label{fig:EPA_900_2300_cl_q}
\end{figure}

\noindent Finally, we consider the real-world network, a subgraph of the WWW obtained by exploring pages linking to www.epa.gov~\cite{Batagelj_Mrvar_2006}, which we refer as EPA in the following (see figure~\ref{fig:EPA_900_2300_cl_q}). 
Also in this case we find a power law behavior with $\beta_q = 0.30 < 0.45 = \beta_{cl}$. The plot displays three areas as mentioned above for the other cases and our conclusions are valid also for this real-world network.

%%%%%%%%%%%%%%%%%%%%%%%%%%%%%%%%%%%%%%%%%
%%%%%%%%%%%%%%%%%%%%%%%%%%%%%%%%%%%%%%%%%
\section{Sensitivity of the quantum PageRank algorithm under coordinated attacks in scale-free graphs}
\label{sect_VII}
%%%%%%%%%%%%%%%%%%%%%%%%%%%%%%%%%%%%%%%%%
%%%%%%%%%%%%%%%%%%%%%%%%%%%%%%%%%%%%%%%%%

In this section, we aim to study of the sensitivity of the quantum PageRank algorithm under coordinated attacks. More precisely, we ask how much the list of quantum PageRanks of an $N$-node graph changes \textit{as a whole} if the $n$ most important nodes (hubs) of a network fail (e.g. due to a hacker attack) and the quantum quantum PageRank algorithm is run on the remaining $(N-n)$ - node graph. Motivated by the fact that the real-world internet belongs to the class of scale-free networks, we focus in our study on the scalefree networks of mesoscopic size (graphs of 16 and 32 nodes).

Operationally, we proceed in our numerical study as follows: 

(i) First, we determine for the initial $N$-node graph the quantum PageRank values and the ordered list of nodes according to the quantum PageRank algorithm. 

(ii) Next, we take out the most important node (main hub) from the graph by eliminating the node itself as well as all its in- and outgoing links from and to the other $(N-1)$ nodes. On this reduced graph (with corresponding modified connectivity matrix $C'$) we carry out the quantum PageRank algorithm to determine of the modified list of quantum PageRanks, with a possibly modified order list of $N-1$ nodes. We note that on this modified graph the algorithm differs quantitatively from the one run on the complete $N$-node network, as the Hilbert space, the initial state and the underlying coherent dynamics are different due to the modified connectivity matrix $C'$ and resulting modified Google matrix $G'$ (see Sec.~\ref{subsect_intro}). 

\begin{figure}
\includegraphics[keepaspectratio=true,width=.82\linewidth]{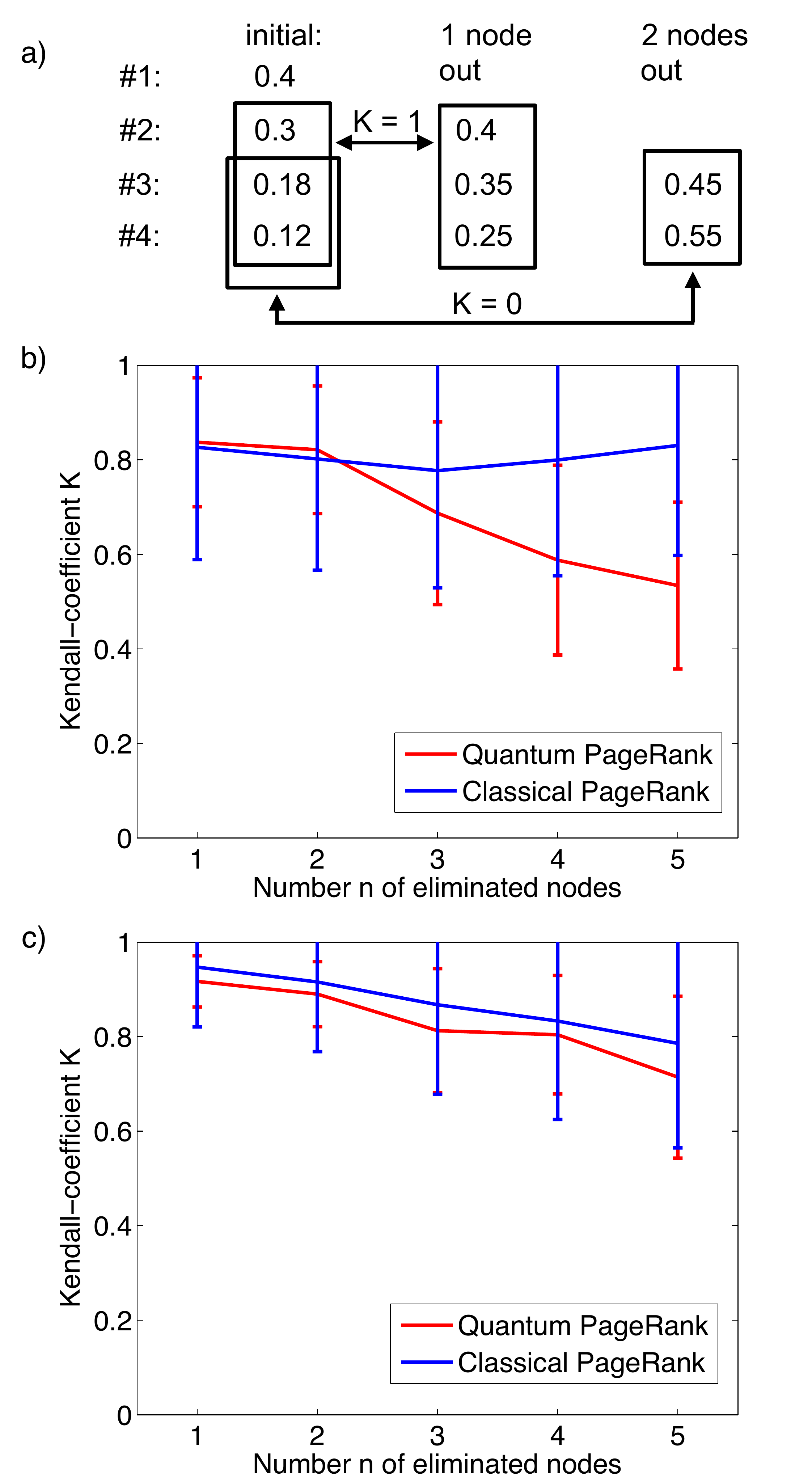}
\caption{
\label{fig:Sensitivity}
(Color online) Numerical study of the sensitivity of the quantum and the classical PageRank algorithm under coordinated attacks in scale-free graphs. a) Conceptual schematics illustrating the comparison of the PageRank list corresponding to networks, where the most important node(s) are taken out, with the PageRank list corresponding to the complete graph with all nodes intact, by means of Kendall's coefficient $K$. b) Numerical results for graphs of 16 and 32 nodes, where up to 5 of the most relevant nodes have been taken out. The data has been obtained by averaging over 100 random scalefree networks, which have been generated with networkX. Error bars correspond to one standard deviation. 
}
\end{figure}

(iii) We compare the ordered list of ($N-1$) nodes according to the quantum PageRank values with the original list for the $N$-node graph, where the most important node is taken out (see Fig.~\ref{fig:Sensitivity}a). To quantify the overall difference between these two lists of $N-1$ elements, we employ Kendall's coefficient \cite{Kendall39}. This function returns for lists in which the \textit{order} of all elements is the same (irrespective of the individual values associated to each element of the list), zero for lists whose order of elements is exactly the opposite, and values in between for lists where the order of elements partially differs. 

This procedure of steps (i) to (iii) can be iterated to take out subsequently the $n$ most important nodes according to the initial quantum PageRank list of the $N$-node graph. The resulting quantum PageRank list of the reduced $(N-n)$ - node graph is then compared to the initial list (not including the $n$ most important nodes). 

To compare the sensitivity of the quantum PageRank algorithm with the classical PageRank algorithm, we perform the same type of coordinated attacks in the classical scenario, i.e., we analyze how the ordered list of classical PageRank values changes when the $n$ most important (according to the \textit{classical} PageRank protocol) nodes of the graph are taken out and the \textit{classical} PageRank algorithm is run to determine the importance of nodes in the reduced $(N-n)$ graphs.

The results are shown in Fig.~\ref{fig:Sensitivity}b and c. We find that when the most important nodes (hubs) are attacked and fail, and the PageRank is recalculated for the reduced graphs, the order of the importance of the remaining nodes changes with respect to the list of the initial complete graph both in the classical and in the quantum case. In the quantum PageRank algorithm, attacks on hubs have a stronger effect than for the classical algorithm. This behavior can be understood by the fact that whereas for the classical case there is a large degeneracy of importance values of nodes of low PageRank, quantum fluctuations partially lift this degeneracy -- see discussion in Sec.~\ref{sect_II} and the insets in Fig.~\ref{fig:MM_EPA_900_2300_Class_PR_vs_QPR}. Thus, when hubs of the network are attacked and fail, the order of less important nodes -- whose importance values slightly differ, can truly change in the quantum case, whereas the degeneracy of a larger number of nodes persists in the classical case. In other works, the increased capability of the quantum algorithm to resolve more structural details of the directed graphs, comes at the cost of an increased sensitivity to structural changes of the network.

%%%%%%%%%%%%%%%%%%%%%%%%%%%%%%%%%%%%%%%%%
%%%%%%%%%%%%%%%%%%%%%%%%%%%%%%%%%%%%%%%%%
\section{Conclusions}
\label{sect_conclusions}
%%%%%%%%%%%%%%%%%%%%%%%%%%%%%%%%%%%%%%%%%
%%%%%%%%%%%%%%%%%%%%%%%%%%%%%%%%%%%%%%%%%

In this paper we have studied the behavior of the quantum PageRank algorithm, developed in Ref.~\cite{gdpmamd2011}, when applied to complex networks. We have found that the quantum algorithm is able to clearly distinguish the structure of the underlying network. More specifically, the quantum PageRank's behavior is distinctive for the three classes of complex networks studied in this work: scale-free networks, graphs of Erd\"os-R\'enyi type and hierarchical networks. 
In particular, we have observed that the quantum algorithm when applied on scale-free networks is able to highlight the structure of the secondary hubs and to resolve the degeneracy in importance of the low lying part of the list of rankings, which represents a typical shortcoming of the classical PageRank algorithm. Although best suited for scale-free graphs the quantum PageRank is also able to univocally uncover whether graphs lie in the Erd\"os-R\'enyi class. Applied to hierarchical graphs 
the algorithm has the capability to better reveal the hierarchy of levels, of which the graph is composed, and to highlight the connectivity structure within every hierarchy layer better than its classical PageRank counterpart. 

Regarding the quantum PageRank algorithm as a directed quantum walk, we have studied the localization properties of the quantum walker in the quantum protocol. By an analysis of the Inverse Participation Ratio (IPR), 
we have observed localization of the quantum walker in the case of the quantum PageRank applied to scale-free networks under standard conditions (damping parameter $\alpha = 0.85 $).
This finding is consistent the ability of the quantum algorithm to highlight hubs of the network. In contrast, for Erd\"os-R\'enyi graphs delocalization was found, which is in accordance with the absence of hubs for this class of networks.

Furthermore, we have analyzed the robustness of the quantum algorithm with respect to variations of the damping parameter. We find a very high degree of robustness as compared to the classical PageRank protocol, which indicates that the value of this parameter, whose choice is to some extent arbitrary, turns out to be not crucial for the quantum algorithm to work reliably.

\noindent Furthermore, we have found that the distribution of importance values of quantum PageRanks over  the nodes of scale-free networks follow - as has been previously found for the PageRank - also a power law behavior.  However, the corresponding scaling exponent is for the quantum protocol smaller than in the classical case, indicating a smoother ranking of nodes. In contrast to the classical algorithm, in the quantum protocol the hubs of the graphs do not concentrate the whole importance and the algorithm lifts the degeneracy of the large set of nodes with low importance values. This increased ranking capability comes at the cost of being more sensitive to structural changes to the network such as coordinate attacks on hubs. 

\noindent Remarkably, the described characteristics of the quantum PageRank even persist if the algorithm is applied to real-world networks. We have studied and successfully tested the performance of the algorithm by applying it to a real-world network, originating from the hyperlink structure of www.epa.org \cite{Batagelj_Mrvar_2006}, thereby showing that its intriguing properties are not restricted to an application of the algorithm to artificially, numerically grown networks.

The classical PageRank algorithm has been the subject of exact studies yielding analytical results, and other interesting studies~\cite{Boettcher_2013,Venegas-Andraca,Whitfield}. 
It would be nice if the quantum algorithm can yield also exact analytical results and for this the class of hierarchical networks is a good candidate.

A related subject of current study is the entanglement properties of quantum complex networks~\cite{Cuquet_Calsamiglia_2012,percolation07,calsamiglia09}. 
A different line of studies has pursued the application of the quantum adiabatic algorithm to the classical PageRank algorithm~\cite{Lidar11}.

In future work it will be interesting to analyze in more detail the impact of random failures of nodes in large networks of differing topology. From an algorithmic point of view, it is an interesting task to develop a dissipative version of this algorithm and to understand its performance and robustness properties in such scenario. Dissipation has already been considered as an element with respect to some aspect of the algorithm  \cite{Garnerone2012,zueco2012}, but the development of a truly dissipative version in the spirit of dissipative quantum algorithms and computation \cite{Verstraete2009,Diehl2009} remains an open question. Furthermore the growing field of complex quantum networks would benefit from a version of the algorithm that is able to rank nodes in the more general case where qubits are located at the nodes of the network. An important question in this scenario is whether an algorithm based on a multi-particle quantum walk \cite{Childs2013,Childs2009} is needed in this context, or if there exists for this task an efficiently simulatable algorithm that belonging to the computational complexity class $P$.

\noindent {\em Acknowledgements.} The authors acknowledge the Centro de Supercomputacion y Visualizacion de Madrid (CeSViMa) for CPU time on the Magerit2 cluster. This work has been supported by the
Spanish MINECO grants and the European Regional Development Fund
under projects  FIS2012-33152, MTM2011-28800-C02-01, CAM research
consortium QUITEMAD S2009-ESP-1594, European
Commission PICC: FP7 2007-2013, Grant No. 249958 and UCM-BS grant GICC-910758.

%%%%%%%%%%%%%%%%%%%%%%%%%%%%%%%%%%%%%%%%%
%%%%%%%%%%%%%%%%%%%%%%%%%%%%%%%%%%%%%%%%%
\section*{Appendix}
\label{Appendix}
%%%%%%%%%%%%%%%%%%%%%%%%%%%%%%%%%%%%%%%%%
%%%%%%%%%%%%%%%%%%%%%%%%%%%%%%%%%%%%%%%%%
\appendix*{

From eq.~\eqref{eq:recap_I_q_alpha_dep}:
\begin{equation}
\langle I_q\rangle (P_i, \alpha) = \left\langle \mathrm{Tr}_1 \left( \mathrm{Tr}_2 \left[  \rho^{12}_\alpha(2t) M^{(2)}_i \right]\right)\right\rangle_t
\label{eq:recap_I_q_alpha_dep_appendix}
\end{equation}
where for bookkeeping purposes in the derivations that follow it has been made explicit to which spaces the density matrix refers to. 
Using the fact that trace preserving quantum operations are contractive. That is:
\begin{equation}
 D\left(\rho^{12}_\alpha(2t) ,\rho^{12}_{\alpha^\prime}(2t) \right) \ge D\left(\mathrm{Tr}_1\rho^{12}_\alpha(2t) ,\mathrm{Tr}_1\rho^{12}_{\alpha^\prime}(2t) \right) \,
\label{eq:trace_dist_contractive_partial_trace}
\end{equation}
and making use of another property, namely that the trace distance of two states is attainable as the maximum over all strong (i.e. PVM) measurement outcomes of the difference of the two states:
\begin{equation}
 D \left(  \rho ,  \sigma   \right)   =    
			\max_{ M }  \mathrm{Tr} 
			\left[ 
 				M \left(  
					 \rho  - \sigma  
  			 	\right) 
 			\right] \,
\label{eq:trace_dist_PVM_prop_gen}
\end{equation}
that, in our case, specializes to:
\begin{align}
 D \left( \mathrm{Tr}_1 \rho^{12}_\alpha(2t) , \mathrm{Tr}_1 \rho^{12}_{\alpha^\prime}(2t) 
 	\right)   = \nonumber \\ =
			\max_{ M^{(2)} }  \mathrm{Tr}_2 
			\left[ 
 				M^{(2)} \left(  
					 \mathrm{Tr}_1\rho^{12}_\alpha(2t)  -\mathrm{Tr}_1\rho^{12}_{\alpha^\prime}(2t)  
  			 	\right) 
 			\right] 
\label{eq:trace_dist_PVM_prop_spec}
\end{align}
we obtain Finally it is useful to measure the average distance as we are interested in the stability of the average quantum PageRank: 
\begin{align}
  \left\langle D \left( \mathrm{Tr}_1 \rho^{12}_\alpha(2t) , \mathrm{Tr}_1 \rho^{12}_{\alpha^\prime}(2t) 
 	\right)  \right\rangle_t = \nonumber \\ =
			 \left\langle \max_{ M^{(2)} }  \mathrm{Tr}_2 
			\left[ 
 				M^{(2)} \left(  
					 \mathrm{Tr}_1\rho^{12}_\alpha(2t)  -\mathrm{Tr}_1\rho^{12}_{\alpha^\prime}(2t)  
  			 	\right) 
 			\right]  \right\rangle_t
\label{eq:trace_dist_PVM_prop_spec_mean}
\end{align}
eq.~\eqref{eq:trace_dist_PVM_prop_spec_mean} can be rewritten as:
\begin{align}
\left\langle D \left( \mathrm{Tr}_1 \rho^{12}_\alpha(2t) , \mathrm{Tr}_1 \rho^{12}_{\alpha^\prime}(2t) 
 	\right)  \right\rangle_t  
	= \nonumber \\ =
	 \max_{ M^{(2)} }  \left\langle
					\left[ 
						  \mathrm{Tr}_1 \mathrm{Tr}_2 M^{(2)}  \rho^{12}_\alpha(2t)  -\mathrm{Tr}_1 \mathrm{Tr}_2  M^{(2)}  \rho^{12}_{\alpha^\prime}(2t)  
  			 		\right] 
				    \right\rangle_t 
\label{eq:trace_dist_PVM_prop_spec_mean_2}
\end{align}
which in our case is clearly stated as:
\begin{align}
  \left\langle D \left( \mathrm{Tr}_1 \rho^{12}_\alpha(2t) , \mathrm{Tr}_1 \rho^{12}_{\alpha^\prime}(2t) 
 	\right)  \right\rangle_t  
	= \nonumber \\ =
	 \max_i \left|
	  \langle I_q(P_i,m,\alpha) \rangle -\langle I_q(P_i,m,\alpha^\prime) \rangle 
	  \right|
\label{eq:trace_dist_stability_Av_QPR_appendix}
\end{align}
}
Where we can take the absolute value because the distance between two states is a nonnegative number.

%%%%%%%%%%%%%%%%%%%%%%%%%%%%%%%%%%%%%%%%%%%%%%%%%%%%%%%%%%%%%%%%%%%%%%%%%%%%%%

\bibliography{bibliography}

\begin{thebibliography}{99}
\bibitem{footnote_fidelity}
This can be seen from a general formula that relates the two measures of distance (see e.g. \cite{NC} 
chap. 9): $1- F(\sigma , \rho )  \le D(\sigma , \rho ) \le \sqrt{ 1 - F(\sigma , \rho )^2} $
\end{thebibliography}

\begin{comment}
%\begin{references}

\end{comment}

\end{document}